%% file: main.tex
\documentclass[lettersize,journal]{IEEEtran}
\IEEEpubidadjcol

\input{includes/includes}

\input{includes/macros}

\input{includes/pygments}

\begin{document}

\title{Automated Code Editing with Search-Generate-Modify}

\author{Changshu Liu, Pelin Cetin$^*$, Yogesh Patodia$^*$, Baishakhi Ray, Saikat Chakraborty, Yangruibo Ding
\thanks{Changshu Liu, Pelin Cetin, Yogesh Patodia, Baishakhi Ray and Yangruibo Ding are affiliated with
Department of Computer Science, Columbia University, New York, NY
USA.
\\
Email: \{cl4062@, pc2807@, yp2607@, rayb@cs., yrbding@cs.\}columbia.edu}
\thanks{Saikat Chakraborty is affiliated with Microsoft Research
Redmond, WA, USA.
\\
Email: saikatc@microsoft.com
}
\thanks{Manuscript received May, 2023; revised Feb, 2024.}}

\markboth{Journal of \LaTeX\ Class Files,~Vol.~14, No.~8, August~2021}%
{Shell \MakeLowercase{\textit{et al.}}: A Sample Article Using IEEEtran.cls for IEEE Journals}


\maketitle

\input{body/0.abstract.tex}


\input{body/1.introduction}

\input{body/2.background.tex}
\input{body/3.approach.tex}
\input{body/3a.patch_search.tex}
\input{body/3b.program_repair.tex}

\input{body/4.experiment.tex}
\input{body/5a.results_and_analysis_pipeline}
\input{body/5e.results_and_analysis_patch_search}
\input{body/5d.results_and_analysis_levenshtein}
\input{body/5b.results_and_analysis_bug_fixing}

\input{body/6.related_work.tex}

\input{body/7.threats.tex}
\input{body/7.5.future_work.tex}
\input{body/8.conclusion.tex}


\bibliographystyle{IEEEtran}
\bibliography{main}

\end{document}

%% file: includes/includes.tex
\usepackage{hyperref}
\usepackage{boldline}
\usepackage{color,colortbl}
\usepackage{bigstrut}
\usepackage[ruled, linesnumbered]{algorithm2e}

\usepackage{amsmath,amsfonts}
\usepackage{footnote}
\usepackage{csquotes}
\usepackage{amsmath}
\usepackage{array}
\usepackage{algorithmic}
\usepackage{balance}
\usepackage{blindtext}
\usepackage{booktabs}
\usepackage{hyperref}
\usepackage[noabbrev]{cleveref}
\usepackage{comment}
\usepackage{enumitem}
\usepackage{eqparbox}
\usepackage{fancybox}
\usepackage{fancyvrb}
\usepackage{flushend}
\usepackage{framed}
\usepackage{graphicx}
\usepackage{ifthen}
\usepackage{listings}
\usepackage{makecell}
\usepackage{mathrsfs}
\usepackage{mdwmath}
\usepackage{mdwtab}
\usepackage{multirow}
\usepackage{pifont}
\usepackage{stfloats}
\usepackage{textcomp}
\usepackage{url}
\usepackage{xspace}
\usepackage[normalem]{ulem}
\usepackage[framemethod=TikZ]{mdframed}
\usepackage{caption}
\usepackage{subcaption}
\usepackage{tabularx}
\usepackage{tcolorbox}
\tcbuselibrary{listings,skins}
\usepackage{mathtools}
\usepackage{blindtext}
\usepackage{lstautogobble}
\DeclareGraphicsExtensions{.pdf,.jpeg,.png}
\graphicspath{{figures/}}

\usepackage[sort&compress, numbers]{natbib}

\usepackage{tcolorbox}

\newtcbox{\mybox}{
  colback=white,
  on line,
}

\newcommand{\rom}[1]{\uppercase\expandafter{\romannumeral #1\relax}}

\newcommand{\etal}{\hbox{\emph{et al.}}\xspace}
\newcommand{\eg}{\hbox{\emph{e.g.,}}\xspace}
\newcommand{\ie}{\hbox{\emph{i.e.,}}\xspace}

\newtheorem{theorem}{Theorem}[section]

\newtheorem{definition}[theorem]{Definition}

\newenvironment{examplebox}{\par\begingroup
  \setlength{\fboxsep}{3pt}\findlength
  \setbox0=\vbox\bgroup\noindent
  \hsize=0.90\linewidth
  \begin{minipage}{0.90\linewidth}\normalsize}
    {\end{minipage}\egroup
    \textcolor{gray20}{\fboxsep1.2pt\fbox
     {\fboxsep5pt\colorbox{gray05}{\normalcolor\box0}}}
    \endgroup\par\noindent
    \normalcolor\ignorespacesafterend}

\newcounter{RQACounter}

\usetikzlibrary{shadows}
\usepackage{graphics}
\newmdenv[
    tikzsetting= {fill=blueish},
    skipabove=0.33em,
    skipbelow=0.33em,
    linewidth=1pt,
    innerleftmargin=4pt,
    innerrightmargin=4pt,
    innertopmargin=2pt,
    innerbottommargin=2pt,
    linecolor=gray95,
    roundcorner=2pt, 
    shadow=true,
    shadowsize=4pt,
    shadowcolor=gray95
]{questionbox}

\newmdenv[
    tikzsetting= {fill=greenish},
    skipabove=0.33em,
    skipbelow=0.33em,
    linewidth=1pt,
    innerleftmargin=4pt,
    innerrightmargin=4pt,
    innertopmargin=2pt,
    innerbottommargin=2pt,
    linecolor=gray95,
    roundcorner=2pt, 
    shadow=true,
    shadowsize=4pt,
    shadowcolor=gray95
]{answerbox}

\newmdenv[
    skipabove=0.33em,
    skipbelow=0.33em,
    innerleftmargin=4pt,
    innerrightmargin=4pt,
    innertopmargin=2pt,
    innerbottommargin=2pt,
]{lessonbox}

\usepackage{tikz}

\def\checkmark{\tikz\fill[scale=0.4](0,.35) -- (.25,0) -- (1,.7) -- (.25,.15) -- cycle;} 

\newenvironment{lesson}
{
    \begin{lessonbox}
}
{\end{lessonbox}}

\newenvironment{result}
{\begin{answerbox}}
{\end{answerbox}}

\newenvironment{question}
{\begin{questionbox}}
{\end{questionbox}}

\definecolor{javared}{rgb}{0.6,0,0} 
\definecolor{javagreen}{rgb}{0.25,0.5,0.35} 
\definecolor{javapurple}{rgb}{0.5,0,0.35} 
\definecolor{javadocblue}{rgb}{0.25,0.35,0.75} 

\lstdefinestyle{basejava}{
  language=java,
  showstringspaces=false,
  basicstyle=\small\ttfamily,
  keywordstyle=\bfseries\color{javapurple},
  commentstyle=\itshape\blue,
  identifierstyle=\blue,
  frame=none,
  backgroundcolor=\color{white},
}

\lstdefinestyle{CustomJava}{
  belowcaptionskip=\baselineskip,
  breaklines=true,
  xleftmargin=\parindent,
  language=java,
  showstringspaces=false,
  basicstyle=\scriptsize\ttfamily,
  keywordstyle=\bfseries\color{javapurple},
  commentstyle=\itshape\blue,
  identifierstyle=\blue,
  belowskip=1pt,
  numbers=left,
  gobble=0
}

\lstdefinestyle{CustomJavaWoNumbers}{
  belowcaptionskip=0.5\baselineskip,
  breaklines=true,
  xleftmargin=\parindent,
  language=java,
  showstringspaces=false,
  basicstyle=\scriptsize\ttfamily,
  keywordstyle=\bfseries\color{javapurple},
  commentstyle=\itshape\blue,
  identifierstyle=\blue,
  belowskip=0.5pt,
  numbers=none,
  gobble=0
}

\lstdefinestyle{codit}{
  belowcaptionskip=\baselineskip,
  breaklines=true,
  language=java,
  showstringspaces=false,
  basicstyle=\scriptsize\ttfamily,
  keywordstyle=\bfseries\color{javapurple},
  commentstyle=\itshape\blue,
  identifierstyle=\blue,
}



%% file: includes/macros.tex
\xspace%

\newcommand{\Comment}[1]{}
\newcommand{\todo}[1]{#1}

\newcommand{\edit}[1]{\todo{{\color{black} #1}}}

\newcommand{\tool}{\textsc{SarGaM}\xspace}

\newcommand{\old}{$v^0$\xspace}
\newcommand{\new}{$v^1$\xspace}
\newcommand{\query}{$v^0_q$\xspace}

\newtcbox{\inlinebox}[1][]{enhanced,
 box align=base,
 nobeforeafter,
 colback=blueish,
 size=small,
 left=0pt,
 right=0pt,
 boxsep=2pt,
 #1}

\newcommand{\sdata}{{$B2F_{s}$}\xspace}
\newcommand{\mdata}{{$B2F_{m}$}\xspace}
\newcommand{\dforj}{{Defects4J$_{1.2}$}\xspace}
\newcommand{\dforjj}{{Defects4J$_{2.0}$}\xspace}

\renewcommand{\cref}[1]{\Cref{#1}}

%% file: includes/pygments.tex
\newcommand\red[1]{\textcolor[rgb]{1.00,0.00,0.00}{#1}}

\newcommand\blue[1]{\textcolor[rgb]{0.00,0.00,1.00}{{#1}}}

\newcommand\purple[1]{\textcolor[rgb]{0.22,0.0,0.44}{#1}}

\definecolor{blueish}{RGB}{250, 250, 255}
\definecolor{greenish}{RGB}{250, 255, 250}
\definecolor{redish}{RGB}{255, 200, 200}

\definecolor{highlight}{RGB}{175, 255, 100}
\definecolor{gray01}{gray}{.98}
\definecolor{gray05}{gray}{0.95}
\definecolor{gray08}{gray}{0.92}
\definecolor{gray10}{gray}{0.90}
\definecolor{gray12}{gray}{0.88}
\definecolor{gray15}{gray}{0.85}
\definecolor{gray18}{gray}{0.82}
\definecolor{gray20}{gray}{0.80}
\definecolor{gray25}{gray}{0.75}
\definecolor{gray30}{gray}{0.70}
\definecolor{gray35}{gray}{0.65}
\definecolor{gray40}{gray}{0.60}
\definecolor{gray45}{gray}{0.55}
\definecolor{gray50}{gray}{0.50}
\definecolor{gray55}{gray}{0.45}
\definecolor{gray60}{gray}{0.40}
\definecolor{gray65}{gray}{0.35}
\definecolor{gray70}{gray}{0.30}
\definecolor{gray75}{gray}{0.25}
\definecolor{gray80}{gray}{0.20}
\definecolor{gray85}{gray}{0.15}
\definecolor{gray90}{gray}{0.10}
\definecolor{gray95}{gray}{0.05}

%% file: body/0.abstract.tex
\begin{abstract}
Code editing is essential in evolving software development. In literature, several automated code editing tools are proposed, which leverage Information Retrieval-based techniques and Machine Learning-based code generation and code editing models. Each technique comes with its own promises and perils, and for this reason, they are often used together to complement their strengths and compensate for their weaknesses. 
This paper proposes a hybrid approach to better synthesize code edits by leveraging the power of code search, generation, and modification. 

Our key observation is that a patch that is obtained by search \& retrieval, even if incorrect, can provide helpful guidance to a code generation model. 
However, a retrieval-guided patch produced by a code generation model can still be a few tokens off from the intended patch. Such generated patches can be slightly modified to create the intended patches. We developed a novel tool to solve this challenge: \tool, which is designed to follow a real developer's code editing behavior. Given an original code version, the developer may {\em search} for the related patches, {\em generate} or write the code, and then {\em modify} the generated code to adapt it to the right context. Our evaluation of \tool on edit generation shows superior performance w.r.t. the current state-of-the-art techniques. \tool also shows its effectiveness on automated program repair tasks. 
\end{abstract}

%% file: body/1.introduction.tex
\section{Introduction}
\label{sec:intro}

In a rapidly-evolving software development environment, developers often edit code to fix bugs, add new features, or optimize performance. This process can be complex and requires a deep understanding of the underlying programming language, as well as an expertise in the relevant domain. To facilitate code editing, developers often search existing codebases~\cite{ray2014uniqueness, nguyen2013study, gharehyazie2017some} or online resources~\cite{rahman2018evaluating} for relevant code, and may also leverage automated code generation tools such as GitHub Copilot\footnote{\href{https://github.com/features/copilot/}{https://github.com/features/copilot/}\label{copilot}}. However, the search results~\cite{zhang2018code, ray2013detecting} or generated code may not always be ideal, necessitating developers to customize them for the given situation~\cite{barke2022grounded}. Therefore, developers may have to further modify the generated code to achieve the desired outcome.

In the past, various tools and techniques have been proposed to reduce the manual effort required for code editing~\cite{boshernitsan2007aligning,robbes2008example,tufano2019learning,chakraborty2020codit}. They can be broadly classified into three different categories: Search \& Retrieve, Generate, and Modify.

\noindent
\textit{Search \& Retrieve.} This is a popular approach to suggest edits that were previously applied to similar code contexts~\cite{ray2014uniqueness,nguyen2013study,meng2013lase,meng2011systematic}. However, 
each retrieval-based technique relies on its perceived definition of code similarity (\eg token, tree, or graph-based similarity) and fails to generate edits with a slight variation of that definition. As a result, these methods tend to have limited applicability to diverse code editing contexts.

\noindent
\textit{Generate.} In recent years, the most promising approach is perhaps the Large Language Model (LLM)-based code generation models where code is generated based on developers' intent and surrounding code context. 
For instance, open-source code-specific LLMs such as PLBART~\cite{ahmad2021unified}, CodeGPT-2~\cite{CodeXGLUE}, CodeT5~\cite{wang2021codet5}, and NatGen~\cite{chakraborty2022natgen} have shown significant potential in code generation. Additionally, industry-scale LLMs like GPT-3~\cite{brown2020language} and Codex~\cite{chen2021evaluating} have gained widespread popularity for generating source code and are used as the backbone of commercial code generation software such as GitHub Copilot$^{\text{\ref{copilot}}}$.

There is a subtle difference between edit generation and code generation. Developers generate edits by transforming the original code version into a new version, often by deleting and adding lines. Edit generation can thus be thought of as a conditional probability distribution, generating new lines of code by conditioning on the old lines. Existing LLM-based code generation approaches do not capture granular edit operations: which tokens will be deleted, which tokens will be inserted, and in particular, where the new tokens will be inserted.

\noindent
\textit{Modify.} Many previous works~\cite{tarlow2020learning,chen2021plur,connorcan2021,zhang2022coditt5} designed special outputs to represent edit operations. Recently CoditT5~\cite{zhang2022coditt5} proposes an edit-specific LLM where given an original code version, CoditT5 \cite{zhang2022coditt5} first comes up with an edit plan (in terms of deletion, insertion, substitution) and then conditioned on the edit plan, it generates the edits in an auto-regressive manner. CoditT5 \cite{zhang2022coditt5} shows promise in generating edits over vanilla code generation.

The goal of this work is to produce higher-quality code edits by harnessing the power of all three techniques. Each approach offers unique ingredients that can contribute to better edit generation.

\textbf{Our Insight.} Code search can retrieve relevant patches that can provide more guidance to a code generation model, leading to better patch generation. However, most of the time the patches generated this way are off by a few tokens from the intended patch---even random permutations and combinations of the generated tokens could lead to the intended patch~\cite{jain2021jigsaw}. 
A more systematic approach would involve using an edit-generation model that specifically targets the generated tokens that require further modifications such as deletion or insertion.
This allows more focused and precise modifications of the code generated in the previous step and finally outputs the intended patch.


\textbf{Proposed Approach.}
We propose a framework, \tool, that leverages code-search-augmented code generation and modification to generate code edits. \tool emulates the typical code editing practice of a developer where given an edit location and context, she might search for related code, write the retrieved code (\ie generation) to the target location, and modify it to contextualize.
\tool contains three steps: 
(i) Search: An information retrieval-based technique to retrieve candidate patches from a database of previous edits that may fit the edit context, 
(ii) Generation: An off-the-shelf code generation model that takes input from the edit location, edit context, and the retrieved patches, and outputs a token sequence corresponding to the edited code, and 
(iii) Modification: A novel code editing model that slightly modifies the token sequence generated in the previous step and outputs granular edit operations in terms of deleted and inserted tokens.

As opposed to the existing edit-generation models~\cite{zhang2022coditt5} that aim to generate the edit operations directly from the original version, we allow a generation model to initially generate the token sequence and then refine it to produce the final patch.
We observe that a granular edit model generally performs better for generating smaller edits. If a generation model already generates a sufficiently accurate patch, enhancing it with further edits can improve the overall effectiveness of the edit-generation model.

\textbf{Results.}
We evaluate our approach on two tasks: code editing and program repair. 
For code editing, we examine \tool on two different datasets. 
\tool improves $top\ 1$ patch generation accuracy over state-of-the-art patch generation models (PLBART~\cite{ahmad2021unified}, NatGen~\cite{chakraborty2022natgen} and CoditT5~\cite{zhang2022coditt5}) \edit{from 19.76\% to 2.77\%} in different settings.
For program repair, we compare \tool with recent Deep Learning-based  techniques on \dforj , \dforjj, and QuixBugs datasets and report state-of-the-art performance. Additionally, we conduct extensive ablation studies to justify our different design choices. In particular, we investigate three components (search, generate, and modify) individually and prove that \tool can benefit from each one of them. 

In summary, our key contributions are:
\begin{itemize}
    \item We prototype a code editing model, \tool, built on top of 
    off-the-shelf pre-trained code generation models and augmented the
    generation model with code search and code modification. 
    
    \item We propose a new code modification model, which involves generating granular edit operations (\ie deletion and insertion operations at token granularity as opposed to generating token sequences).

    \item We demonstrate \tool's ability to generate patches for general-purpose code edits and bug fixes. Across most of the settings, \tool achieves state-of-the-art performances. We present a detailed ablation study to justify our different design choices.

    \item We release our prototype tool at \\ \href{https://github.com/SarGAMTEAM/SarGAM.git}{https://github.com/SarGAMTEAM/SarGAM.git}.

\end{itemize}

%% file: body/2.background.tex
\begin{figure*}
     \centering
     \begin{subfigure}[b]{0.25\linewidth}
         \centering
         \includegraphics[width=\textwidth]{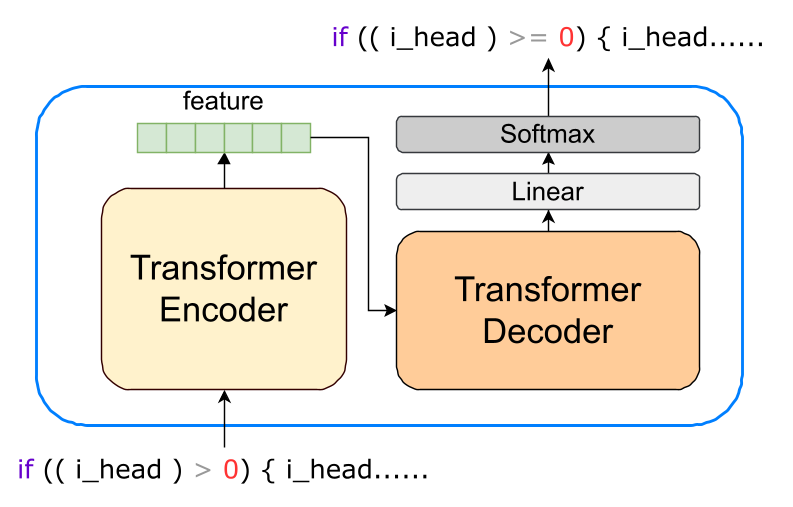}
         \caption{\small{Encoder-Decoder}}
         \label{fig:encoder-decoder}
     \end{subfigure}
     ~~
     \begin{subfigure}[b]{0.25\linewidth}
         \centering
         \includegraphics[width=\textwidth]{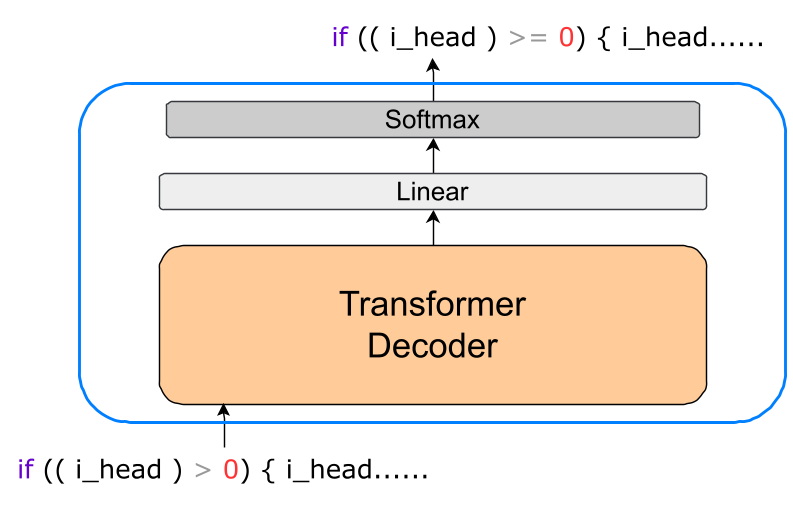}
         \caption{\small{Decoder (Only)}}
         \label{fig:decoder-only}
     \end{subfigure}
     ~~
     \begin{subfigure}[b]{0.25\linewidth}
         \centering
         \includegraphics[width=\textwidth]{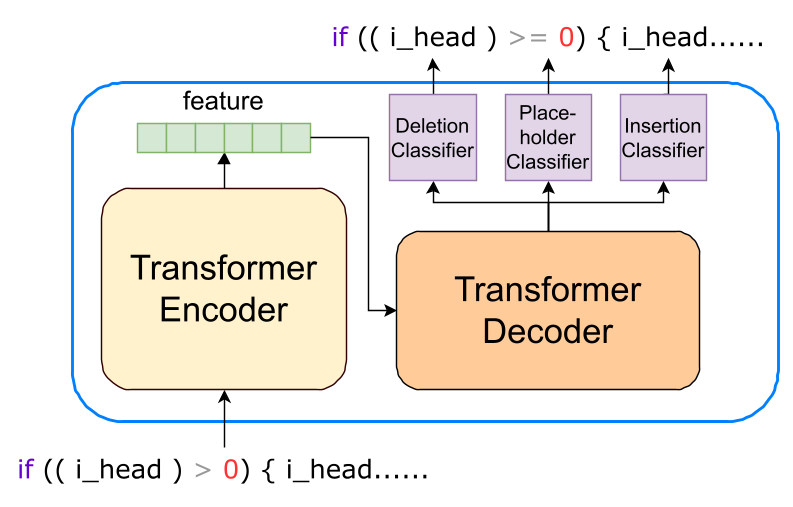}
         \caption{\small{Levenshtein Transformer}}
         \label{fig:lev-transformer}
     \end{subfigure}
        \caption{\small{Different Types of Transformer-based Generative Models}}
        \label{fig:three graphs}
\end{figure*}
\vspace{-10pt}

\section{Background: Code Generation Models}

Machine Learning-based Code Generation has gained significant attention in recent years, where code is generated from a Natural Language (NL) description or code context. 
Different types of Sequence-to-Sequence ($seq2seq$) models play a significant role in achieving this success~\cite{pradel2021neural, ding2020patching}.
The input to a $seq2seq$ model is a sequence of tokens ($X = x_1, x_2, ..., x_n$), and the output is a token sequence ($Y = y_1, y_2, ..., y_m)$, where  the model learns conditional probability distribution $P(Y|X)$. 

Recurrent Neural Networks (RNN) and Long Short Term Memory (LSTM)-based models~\cite{hochreiter1997long} once held a dominant role in code generation~\cite{chen2019sequencer, chakraborty2020codit, 
nam2022predictive,yin2018learning}. RNNs and LSTMs take a piece of code token-by-token in a sequential manner and try to predict the next token conditioned on the immediately preceding tokens in the sequence. The two types of models largely depend on the tokens in close vicinity and tend to suffer from not capturing the long-range dependencies~\cite{zhao2020rnn}.

\subsection{Transformer for Code Generation}
\label{sec:back_tr}


Transformer-based models~\cite{vaswani2017attention} have recently outperformed alternative architectures for code generation due to the introduction of the self-attention mechanism.
Transformers process the entire input token sequence as a complete graph 
\footnote{\href{https://en.wikipedia.org/wiki/Complete_graph}
{https://en.wikipedia.org/wiki/Complete\_graph}}. Each token is a vertex in the graph, and an edge connecting two vertices is the \enquote{attention} between the corresponding tokens. 
The attention is the relative influence of a token to represent other tokens in the sequence. The attention weights signify the importance of a token to make the final prediction for a particular task~\cite{kobayashi2020attention, abnar2020quantifying}. 
The model learns the attention weights depending on the task during the training process. The Transformer also encodes the  relative position of each token in the input sequence  (positional encoding). 

The attention mechanism and positional encoding allow Transformers to catch more long-range dependencies. The self-attention mechanism allows parallel processing of input sequences that leads to significant speedup during training~\cite{vaswani2017attention}. Many previous works 
use Transformers for code generation problems (\eg patching, code editing, and program repair) due to their success 
\cite{jiang2023knod,xia2022less,chakraborty2021multi, chakraborty2022natgen}. 
Transformer-based models roughly fall into two categories: encoder-decoder and decoder-only.

\textbf{Encoder-decoder}. As shown in~\Cref{fig:encoder-decoder}, an encoder-decoder model has a Transformer encoder and an autoregressive Transformer decoder. The encoder is trained to extract features from the input sequence. 
The decoder generates the next token by reasoning about the feature extracted by the Transformer encoder and previously generated tokens.
PLBART \cite{ahmad2021unified}, CodeT5~\cite{wang2021codet5}, and NatGen~\cite{chakraborty2022natgen} are  examples of encoder-decoder models trained on code corpora with denoising pre-training. 
CoditT5~\cite{zhang2022coditt5} further presents a custom pre-trained model 
for code editing tasks using the same architecture as CodeT5 \cite{wang2021codet5}.
MODIT~\cite{chakraborty2021multi}, on the other hand, fine-tunes PLBART \cite{ahmad2021unified} for code editing tasks.
 


\textbf{Decoder-only}. Decoder-only models only have an autoregressive Transformer decoder (shown in Figure \ref{fig:decoder-only}).
Since there is no encoder, decoder-only  transformer is a 
\enquote{generate only} architecture. Such models are pre-trained in an unsupervised way from large corpora to build Generative Pre-trained Models (GPT).
Jiang \etal \cite{jiang2021cure} shows the effectiveness of GPT for the task of source code patching.
Other representative decoder-only code generation models include Ploycoder~\cite{xu2022systematic}, OpenAI's Codex\cite{chen2021evaluating}, GPT-3~\cite{brown2020language}, etc. 
Decoder-only models are suitable for open-ended code generation, where a prompt describing the functionality is passed to the model.

\subsection{Levenshtein Transformer}
\label{sec:back_levt}

The Transformers usually generate outputs from scratch. When there is much overlap between input and output token sequences (\eg automatic text editing  where only a few tokens are changed, keeping most of the text as it is), Transformers tend to suffer~\cite{gu2019levenshtein} to preserve the unchanged tokens. 
Levenshtein Transformers (LevTs)~\cite{gu2019levenshtein} show promises in such cases, as they use basic edit operations such as \emph{insertion} and \emph{deletion} to implement granular sequence transformations.
Levenshtein Distance \cite{levenshtein1966binary} between the ground truth and the output token sequence is measured during training after each deletion or insertion. 
The predicted operation is chosen for the next interaction if the distance reduces.

\Cref{fig:encoder-decoder} and \Cref{fig:lev-transformer} show architectural differences between a Transformer and a LevT. Although both share the same encoder and decoder blocks, the vanilla Transformer uses a  linear layer and softmax upon stacks of decoder layers to predict the next token, while LevT uses three additional classifiers to apply edit operations. In LevT, the output of the last Transformer decoder block (\eg 
$h=\{ h_0, h_1, \cdots, h_n \}$) is passed to following classifiers:

\begin{enumerate}[leftmargin=20pt]
  \item Deletion Classifier: for each token in the sequence, this binary classifier predicts whether it should be deleted(=1) or kept(=0).
  $\pi_\theta^{\mathrm{del}}(h_i) = softmax(W_{del}h_i)$, 
   where $W_{del}$ is the weight matrix of the deletion classifier.
   
  \item Placeholder Classifier: predicts how many {place holders} should be inserted between any consecutive token pairs in the whole sequence.
      $\pi_\theta^{\mathrm{plh}}(<h_i, h_{i+1}>) = softmax(W_{plh}\cdot concat(h_i, h_{i+1}))$
  ,where $W_plh$ is the weight matrix of the placeholder classifier.
  
  \item Insertion Classifier: for each placeholder we inserted in the p revious step, the insertion classifier predicts which token should be inserted in this position: 
      $\pi_\theta^{\mathrm{ins}}(h_i) = softmax(W_{ins}h_i)$
  ,where $W_{ins}$ is the weight matrix of the insertion classifier.
\end{enumerate}


We choose various Transformer-based code generation models to generate patches and use a novel LevT-based edit generation model to further edit the generated patches.

%% file: body/3.approach.tex
\section{\tool Approach}
\label{sec:approach}

\begin{figure}
\centering
  \includegraphics[width=0.30\textwidth]{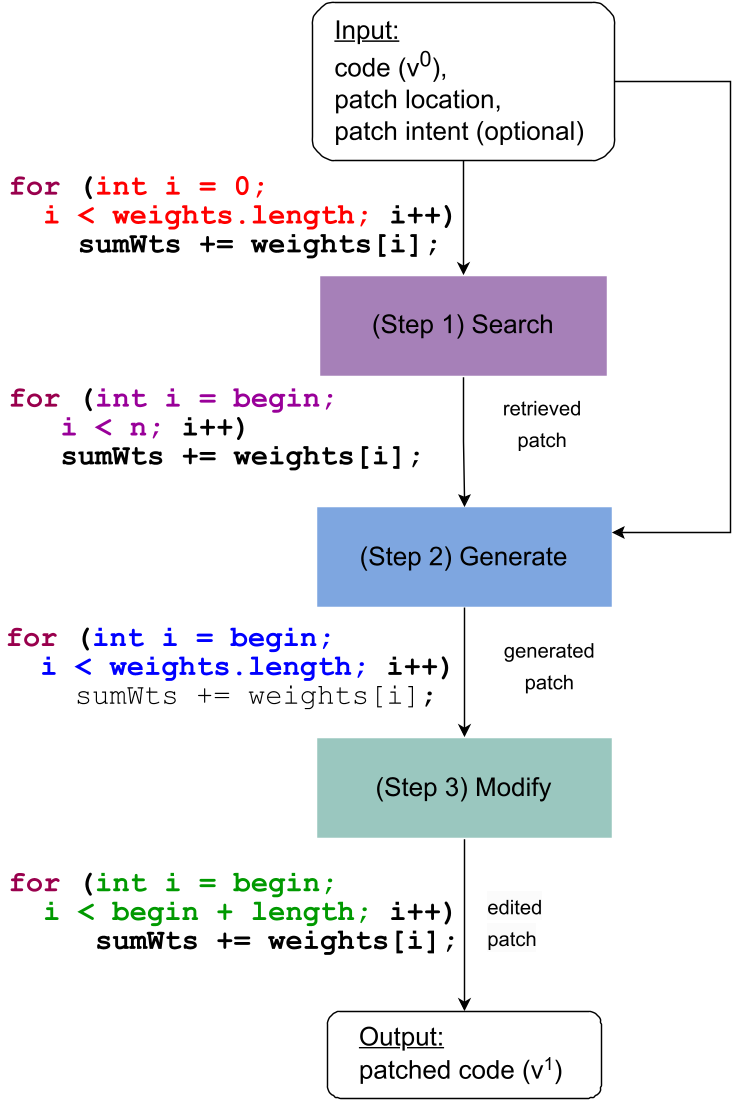}
  \caption{\small{Overview of the \tool Pipeline and a Motivating Example of a bug fixing patch taken from \dforj dataset. {\normalfont Here, inside a for loop, the loop counter initialization and loop condition (\texttt{{ int i=0; i\textless weights.length }}) are buggy and (\texttt{int i=begin; i\textless begin+length}) is the expected fix. After the {\em Search} (Step 1), \tool retrieves a similar patch (\texttt{int i=begin; i\textless n}), the retrieval of \texttt{begin} token benefits {\em Generation} (Step 2). The generated patch is close to the ground truth: (\texttt{int i=begin; i\textless weights.length}), yet not correct. Finally, the {\em Modification} model (Step 3) further modifies the generated patch by deleting \texttt{weights.} and inserting \texttt{begin+}.}}
  }
  \label{fig:overview}
\end{figure}

We introduce \tool, a tool to synthesize source code edits (\ie patches).
A patch is a set of edits to the source code used to update, fix, or improve the code. Throughout the paper, we use code edits and patches interchangeably. More formally, 

\begin{definition}\label{def:3.1}
A program patch, $p:=\Delta(v^0,v^1)$, is the set of syntactic differences that update a program version \old to \new. 
Each change is a token that is either deleted from \old or inserted to \new. 
\end{definition}
We have designed \tool to mimic a real developer's behavior while patching source code.
Given \old, the developer may (i) {\em search} for the related patches, (ii) {\em generate} or write the code, and (iii) further {\em modify} the generated code to adapt it to the right context. 
To this end, we design \tool to have these three steps: Search, Generate, and Modify. An overview of \tool’s workflow is shown in \Cref{fig:overview}.

\subsection{Overview} 

\tool takes the following as input: \old, the exact edit location (can be captured using the current cursor location), and optional edit intent in the form of NL.
\tool then proceeds through the Search, Generate, and Modify steps, ultimately producing the final code version, \new.

\begin{itemize}[leftmargin=*]
    \item 
    \textit{Step 1. Search}: 
    Given the input as a query, \tool searches a database of patches to find similar patches applied previously in a similar context. 
    This step is similar to a search-based patch recommendation engine~\cite{ray2014uniqueness}.
    Each retrieved patch is concatenated with the original input and passed to the next step (see~\Cref{fig:input_format}). 
    In the motivating example in~\Cref{fig:overview}, given the buggy code as query, the search retrieves a similar patch from the code base: \purple{for (int i=begin; i\textless n; i++)}. Although the retrieved patch is not perfect, the introduction of the \textless begin\textgreater\ token facilitates the final result.
    

    \item 
    \textit{Step 2. Generate.} 
    This step takes the search augmented input and outputs a token sequence to generate the patched code. 
    We use off-the-shelf $seq2seq$ models~\cite{ahmad2021unified, chakraborty2021multi, chakraborty2022natgen}, as discussed in~\Cref{sec:back_tr}, to generate code.
    ~\Cref{fig:overview} shows the generation step produces a token sequence \texttt{for (int i=begin; i\textless weights.length; i++)}, which is close to the intended patch. 
    
    
    
    
    \item 
    \textit{Step 3. Modify.} 
     However, the generated patch can still be incorrect, as shown in our running example --- often, they are quite close to the intended edit (\ie low edit distance), nevertheless, incorrect \cite{jain2022jigsaw}.
     Developers still need to modify the generated patch here and there to get the intended output. In this step, we aim to capture such small modifications by explicitly modeling them as deleted and added operations. 
     Our key insight is, as there is a significant overlap between the source and target token sequences, learning granular edit operations can be beneficial.
     In particular, we use LevT, as described in~\Cref{sec:back_levt}, to explicitly model the edits as a sequence of deleted and added tokens. 
     In the case of~\Cref{fig:overview}, this step explicitly deletes {weights.} and adds token sequence {begin+}, resulting in the correct patch. 
\end{itemize}

In this work, we implemented our own Search and Edit Models on top of existing generation models~\cite{ahmad2021unified,chakraborty2021multi,chakraborty2022natgen,zhang2022coditt5}. The rest of this section elaborates each part in detail.

\subsection{Input Processing}
\label{sec:input}



While pre-processing the inputs, following MODIT \cite{chakraborty2021multi}, we create a multi-modal input capturing (i) exact patch location, (ii) patch context, (iii) developers' intent or guidance in the form of natural text. 
Figure \ref{fig:input_format} provides an example input. 
Following some recent code editing and program repair techniques \cite{jiang2021cure,lutellier2020coconut,chakraborty2021multi,zhu2021syntax}, we assume that the developer knows the exact patch location. 
We realize the localization with a tree-based parser based on GumTree \cite{falleri2014fine}.
Patch context is the whole function body where the patch will be applied (including both the context before and after the patch location). 
The third modality (intent) is optional and approximated from the patch commit message. 
We further augment each input with a token sequence retrieved from the search step as discussed below. 
Each modality is separated by {\tt <s>}.

The retrieved patch is inserted after the patch location. We assume that due to the small length of the patch location and retrieved patch, information  from retrieved patch will not be lost during truncation. To ensure that all the models are given the same information, we use the same context window size (512 tokens after tokenization) as PLBART \cite{ahmad2021unified}, CoditT5 \cite{zhang2022coditt5}, and NatGen \cite{chakraborty2022natgen}. Strictly following \cite{chakraborty2021multi} and \cite{zhang2022coditt5}, for samples exceeding the context window size, we simply truncate at the end. 
In the fine-tuning stage, the Transformer model is trying to adaptively capture the relationship between different modalities in order to minimize the training loss. As a result, the model is supposed not to copy too many tokens when the retrieved patch or the commit message is not very similar to the expected patch. These also explain why the generation model still benefits from retrieved patches although they are not even close to the ground truth under some settings in Table \ref{table:averaged_dis}.For code editing tasks, the input of each modality is a list of tokens. 
We tokenize the input with the tokenizer which is compatible with the backbone generation model. 

%% file: body/3a.patch_search.tex
\subsection{Search}


\begin{figure}
    \centering
    \includegraphics[width=0.8\columnwidth]{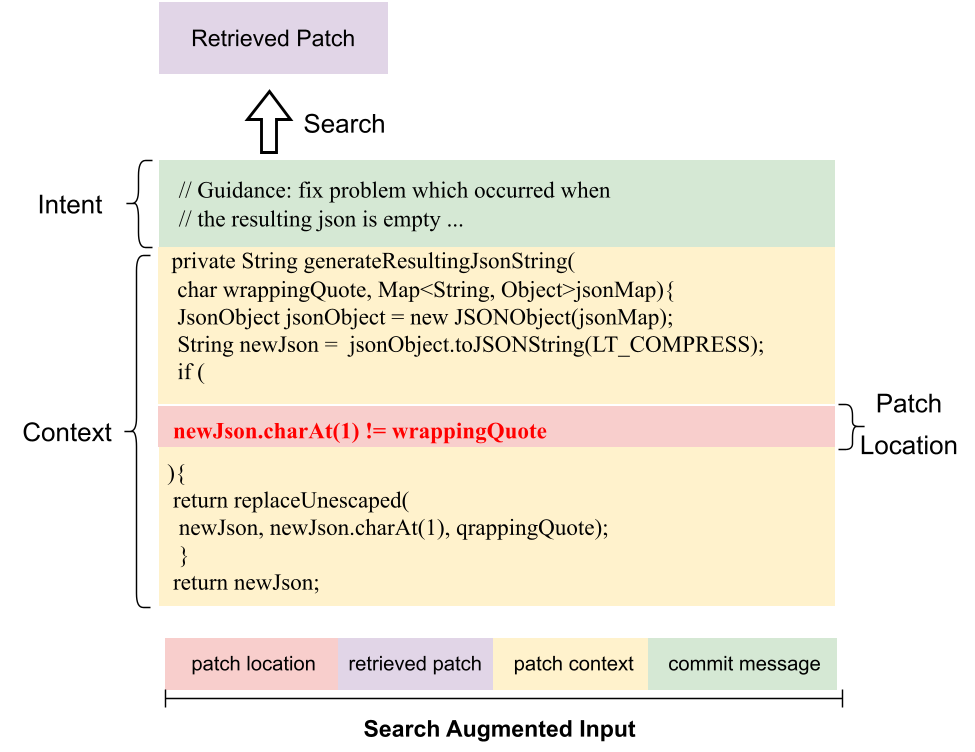}
    \caption{\small{Search-Augmented Input Modalities of \tool}}
    \label{fig:input_format}
\end{figure}

We maintain a database $P$ of previous patches, where each patch can be stored as tuple (\old,\new).
In this step, given the original code version as a query, \query, \tool retrieves the potential edits from the database that were applied before in a similar code context. 
 In particular, \tool utilizes a brute-forced approach: it computes cosine similarities between \query and all the instances (an instance refers a patch along with its corresponding patch location, patch context and commit message) of \old in the database and creates a ranked list based on the similarity scores.
\tool then retrieves $top\ k$ similar $v^0$s and fetches their corresponding patches, $v^1$s. Each retrieved patch is then augmented with the original input, as shown in~\Cref{fig:input_format}.

To ensure the information of all the modalities are passed into our system, for the retrieval model, the window size is 1024. No dimensionality reduction was performed in the search component.

\begin{algorithm}
\footnotesize
\caption{\small{Pseudo Code of Search}}\label{alg:search}
\KwData{
\\1. A query $v^0_q$ as an original code version to be patched, \\ 
2. Patch database $P=\{(v^0_1,v^1_1),..(v^0_i,v^1_i), (v^0_N,v^1_N)\}$, stored with embedding of each $v^0_i$ : $\mathcal{E}(v^0_i)$ \\ 
3. Number of patches to be retrieved $k$;}
\KwResult{Retrieved Patches}
$retrievedP$ = [] \;
\For{p in P}{
  d = Distance($\mathcal{E}(v^0_q)$, $\mathcal{E}(v^0_p)$)\;
  $retrievedP$.append( \{patch:$v^1_p$, distance:d\}) \;
}
Sort $retrievedP$ using distance \;
\Return $retrievedP[:k]$
\end{algorithm}
\vspace{-10pt}

\Cref{alg:search} shows the pseudo-code for our technique.
As inputs, the algorithm takes an original code version that needs to be patched (\query), a database of previous patches $P$, and how many patches we want to retrieve ($top\ k$).
For each original version of a patch $v^0_p$ in the database,  
we compute its edit distance from \query. 
We compute the edit distance in the embedded space to facilitate the computation. Thus, all the original code versions in the patch database are stored in its embedded form $\mathcal{E}(v^0)$, and the query is also embedded. We use PLBART~\cite{ahmad2021unified} to get such embeddings. 
The edit distance is computed using cosine similarity---for any two pieces of embedded code $x$ and $y$, we compute:


\begin{equation}
d = 1-\frac{\mathbf{x} \cdot \mathbf{y}}{\|\mathbf{x}\|\|\mathbf{y}\|}=1-\frac{\sum_{i=1}^N x_i y_i}{\sqrt{\sum_{i=1}^N x_i^2} \sqrt{\sum_{i=1}^N y_i^2}}
\end{equation}

For each candidate \textit{p} in the database, the computed distance along with the retrieved patch ($v^1_p$) is stored in a list (line 4). 
The final list is further sorted in descending order by distance to the query (line 6), and the algorithm returns the $top\ k$ closest entries to the query (line 7). 

Such similarity measurements simulate the situation 
where the developer looks for use cases on the
internet and chooses the problem statement most similar to their
scenario. 

%% file: body/3b.program_repair.tex
\subsection{Generation Model}
Here we use three state-of-the-art edit generation models: PLBART \cite{ahmad2021unified}, CoditT5 \cite{zhang2022coditt5}, and NatGen \cite{chakraborty2022natgen}.
The output of this step is a token sequence corresponding to the generated patch. 
For PLBART\cite{ahmad2021unified} and NatGen\cite{chakraborty2022natgen}, the output formats are identical to the expected patch format and no more post-processing is needed.
However, CoditT5's~\cite{zhang2022coditt5} is an edit generation model; its output sequence is of the format: \emph{Edit Operations} \textless s\textgreater\  \emph{fixed code}. Thus, we further post-process them to create a sequence of tokens corresponding to the generated patch.

\subsection{Modification Model}

\begin{figure}[h]
\centering
    \includegraphics[width=0.7\linewidth]{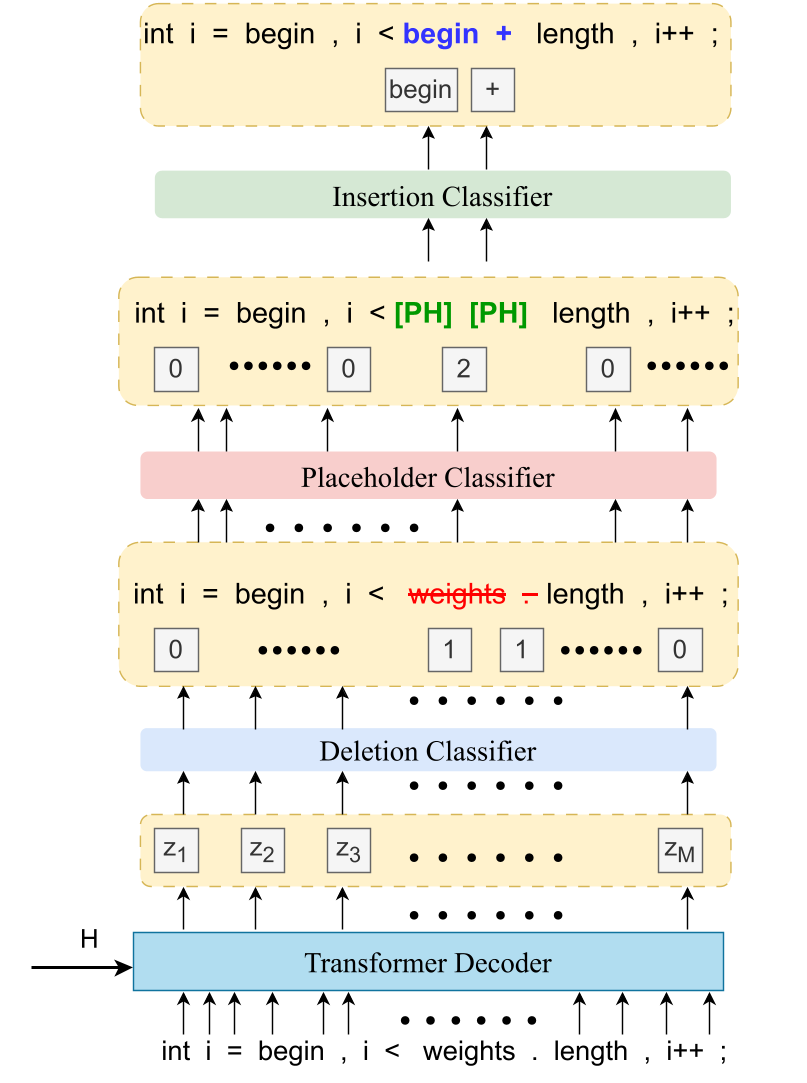}\par
    \centering
    \caption{\small{Example modification steps generated by Levenshtein Transformer corresponding to the motivating example. 
    The encoder takes patch location, context, and optional developer's intent as input and outputs hidden state $H=\{h_1, h_2, \cdots, h_N\}$, where N refers to the length of the input sequence. LevT decoder takes $H$ and patch location, and after some Transformer decoder layers, outputs $(z_1, z_2, \cdots, z_M)$. It is passed to three classifiers (deletion, placeholder, insertion) to perform the edits.}}
    \label{fig:levt_arch}
\end{figure}

Here, a generated code, \eg $v_{gen}$, from the previous step is further modified.
We describe two basic edit operations on $v_{gen}$:
\begin{itemize}
    \item \textbf{delete} token $d$ from $v_{gen}$.
    \item \textbf{insert} token $i$ at location $l$ in $v_{gen}$.
\end{itemize}

Any other code change operation, \eg  replacement, move, etc., can be expressed in terms of delete and insert~\cite{meng2011systematic,meng2013lase}. 
Multiple modifications can further be expressed as a sequence of token deletion and insertion operations, resulting in the final patched code. To capture such insertion-deletion operations, we use LevT, as discussed in~\Cref{sec:back_levt}. ~\Cref{fig:levt_arch} illustrates this step w.r.t. our motivating example (see~\Cref{fig:overview}).


\textbf{Modeling Edits.} Given a token sequence representing $T=(t_1, t_2, ..., t_n)$, 
 the two edit operations, deletion and insertion, are consecutively applied to generate the final output. As discussed in~\Cref{sec:back_levt}, LevT decoder has three classification heads: Insertion, Deletion, and Placeholder. We model the code edit operations using these three classifiers, as follows:

\noindent
\textit{Token Deletion.} LevT reads the input sequence $T$, and for every token $t_i \in T$, the deletion classifier makes the binary decision of 1 (delete the token) or 0 (keep the token). The output of the deletion head is $T'$. ~\Cref{fig:levt_arch} shows that 
the deletion classifier identifies the tokens \texttt{weights} 
and \texttt{.} for deletion (marked in red).

\noindent
\textit{Token Insertion.} On $T'$, the insertion operation is realized in two phases: predicting locations to insert the token and then predicting the tokens to be inserted. First, among all the possible locations where a new token can be inserted, \ie $(t'_i,t'_{i+1}) \in T'$, the Placeholder head of LevT predicts how many placeholders can be added. Next, the insertion head of LevT replaces each placeholder with a  token chosen from the vocabulary. 

For instance, in~\Cref{fig:levt_arch}, the Placeholder Classifier predicts two placeholder positions between tokens \texttt{<} and \texttt{length}, as marked by \texttt{[PH] [PH]} (\ie \texttt{i \textless\ [PH] [PH] length}). 
Next, the Insertion Classifier focuses only on the two placeholders and  predicts \texttt{begin} and \texttt{+} respectively. 
Finally, we get the intended patch \texttt{int i = begin , i < begin + length , i++ ;}.

%% file: body/4.experiment.tex
\section{Experimental Design} 
\subsection{Datasets}

\begin{table}[htpb]
\centering
\caption{\small{Studied Code Editing \& Bug-fixing Datasets}}
\label{table:data}
\scalebox{0.85}{
\setlength\tabcolsep{1.5pt}
\begin{tabular}{l|rrr}
\toprule
\textbf{Dataset}        & \textbf{\#Train} & \textbf{\#Valid} & \textbf{\#Test} \\ \midrule
\textbf{Bug2Fix small} (\sdata)  & 46,628            & 5,828            & 5,831            \\
\textbf{Bug2Fix medium} (\mdata) & 53,324            & 6,542            & 6,538            \\
\textbf{CoCoNuT pre-2006}         & 2,593,572            & 324,196            & -               \\
\textbf{\dforj}  & -                & -               & 75              \\
\textbf{\dforjj}  & -                & -               & 82              \\ 
\textbf{QuixBugs} &-            & -     &40 \\
\bottomrule
\end{tabular}
}
\end{table}
\vspace{-10pt}

~\Cref{table:data} summarizes the dataset we use for our study. 
\subsubsection*{Code Editing Data}
The accuracy of the code editing task of \tool is evaluated by utilizing the Bug2Fix dataset~\cite{tufano2018empirical} similar to \cite{chakraborty2021multi, zhang2022coditt5}  (including $B2F_{s}$ and $B2F_{m}$). 
$B2F_{s}$ contains shorter methods with a maximum token length 50, and $B2F_{m}$ contains longer methods with up to 100 token length.

\subsubsection*{Bug Fixing Data}
The effectiveness of the pipeline is measured with Defects4j \cite{just2014defects4j} and QuixBugs\cite{lin2017quixbugs}. 
The Generate and Edit parts of the pipeline are trained with CoCoNuT pre-2006 \cite{lutellier2020coconut}, which has over two million samples in the training set. Since the bugs in CocoNut pre-2006 are older than the first bug in benchmarks we used, there is no risk of having patched code in the training set. After training, we test 
our pipeline on (1) 75 single-line bugs in \dforj and (2) 85 single line bugs in \dforjj and (3) 40 bugs in QuixBugs.

\subsection{Training}
We trained LevT on 4 GeForce RTX3090 Ti GPUs with a 64,000 tokens batch size, following \cite{gu2019levenshtein}, and applied a dual-policy learning objective, stopping when validation set performance plateaued for 5 consecutive epochs. For code editing, we fine-tuned PLBART, CoditT5, and NatGen, using learning rates of $5e^{-5}$, with batch sizes of 16 for PLBART and 48 for CoditT5 and NatGen, adhering to strategies from relevant literature, and implemented the same early stopping criterion as in LevT training.
\subsection{Baselines}
We fine-tune three large-scale pre-trained language generation models: PLBART \cite{ahmad2021unified}, CoditT5 \cite{zhang2022coditt5} and NatGen \cite{chakraborty2022natgen} on each dataset and consider them as our baselines.
CoditT5\cite{zhang2022coditt5} is an edit-generation model that generates edits in terms of token addition, deletion, and replacement operations. In contrast, NatGen~\cite{chakraborty2022natgen} and PLBART~\cite{ahmad2021unified} are code-generation models that generate a sequence of tokens. Another edit generation model, MODIT~\cite{chakraborty2021multi} studied several information modalities on top of PLBART~\cite{ahmad2021unified}. We use MODIT's \cite{chakraborty2021multi} recommendation to select the input modalities and report results on the different baselines. We compare \tool with the following deep learning-based baselines for the bug-fixing task: CocoNut \cite{lutellier2020coconut}, CURE \cite{jiang2021cure}, KNOD \cite{jiang2023knod}, and AlphaRepair \cite{xia2022less}.



\subsection{Evaluation Metric}
We use accuracy (exact match) 
to evaluate the results on both code editing and program repair. 
When a synthesized patch is exactly identical to the expected patch, we call the synthesized patch the correct one. 
For code editing tasks, we report $top\ 1$ and $top\ 5$ accuracies. Given a retrieval augmented input, we let the code generation model output up to $top\ 5$ patches; modify each of the generated patches once and produce up to 5 final candidate patches. 
 Following \cite{zhang2022coditt5}, we  apply statistical significance testing using bootstrap tests \cite{berg2012empirical}
with confidence level 95\%. The result with the same prefixes (\eg $\alpha$ ) are not significantly different.

In the case of our program repair tool, we generate and evaluate up to the top 1250 patches. We made this choice in consideration of other APR tools, which often evaluate up to the top 5000 patches. We believe that reporting accuracy at the top 1250 is a reasonable and fair choice, particularly because our APR approach includes test cases to validate the generated patches.

%% file: body/5a.results_and_analysis_pipeline.tex
\section{Results and Analysis}
In this section, we empirically evaluate:
\newlist{questions}{enumerate}{2}
\setlist[questions,1]{label=RQ\arabic*.,ref=RQ\arabic*}
\setlist[questions,2]{label=(\alph*),ref=\thequestionsi(\alph*)}

\begin{questions}

    \item How effective is \tool for code editing? 
    \item What are the contributions of different design choices?
    \begin{enumerate}[labelindent=0pt]
        \item Importance of input modalities. 
        \item Effectiveness of a Levenshtein transformer over a vanilla transformer for patch modification. 
     \end{enumerate}
    \item How effective is \tool for automated program repair?
\end{questions}

\subsection{RQ1. \tool for code editing}
\label{rq1}
\subsubsection{Motivation}
Here we investigate the core functionality of \tool, \ie generating code edits. We evaluate it on popular code editing benchmarks \sdata and \mdata.
\subsubsection{Experimental Setup}
We compare \tool's performance with three state-of-the-art pre-trained code generation models that show effectiveness for code editing tasks: 
PLBART \cite{ahmad2021unified}, CoditT5 \cite{zhang2022coditt5}, and NatGen \cite{chakraborty2022natgen}. 
We fine-tune all three pre-trained models on the same dataset ($B2F_{s}$ or $B2F_{m}$). 
While comparing with a code generation model, we incorporate the same model in \tool's pipeline.
In that way, it shows how much \tool can improve compared to the corresponding generation-only setting.

In the search step, we search for similar edits from the training sets of $B2F_{s}$ or $B2F_{m}$. The retrieved patch from the training sets of $B2F_{s}$ or $B2F_{m}$ are added to the input. The generation and edit models are fine-tuned on the search augmented input.
For a given retrieval augmented input, we take $top\ 1$ and $top\ 5$ outputs from the generation step and further modify them to produce the final patches. The reported numbers in~\Cref{table:code_editing} present the accuracy of the final patches. 

\begin{table}[]
\centering
\caption{\small{Exact match of \tool for code editing. Models in () are the off-the-shelf  generative models used by \tool. }}
\label{table:code_editing}
\scalebox{0.9}{
\setlength\tabcolsep{1.5pt}
\begin{tabular}{l|cc|ll}
\hline
\multicolumn{1}{c|}{\multirow{2}{*}{{Tool}}} & \multicolumn{1}{c|}{\textbf{\sdata}} & \textbf{\mdata}                     & \multicolumn{1}{l|}{\textbf{\sdata}} & \textbf{\mdata} \\ \cline{2-5} 
\multicolumn{1}{c|}{}                               & \multicolumn{2}{c|}{{Top1}}                                         & \multicolumn{2}{c}{{Top5}}                      \\ \hline
{PLBART}                                     & 29.99                                & 23.03                               & 47.08                                & 36.51           \\
{\tool (PLABRT)}              & 35.77                                & 27.58                               & 52.43                                & 37.81           \\ \hline
{NatGen}                                     & 36.55                                & $28.53^\alpha$                               & 52.39                                & 42.99           \\
{\tool(NatGen)}               & 38.31                                & $29.32^\alpha$                               & 57.31                                     & \textbf{45.31}  \\ \hline
{CoditT5}                                    & \multicolumn{1}{l}{37.52}            & \multicolumn{1}{l|}{28.33}          & 54.99                                & 42.32           \\
{\tool(CoditT5)}              & \multicolumn{1}{l}{\textbf{39.54}}   & \multicolumn{1}{l|}{\textbf{30.12}} & \textbf{57.46}                       & 44.20           \\ \hline
\end{tabular}
}
\end{table}


\subsubsection{Results}

We find that \tool can always outperform its pure generation counterpart
by a considerable margin. \tool can relatively improve PLBART \cite{ahmad2021unified}, NatGen \cite{chakraborty2022natgen}, and CoditT5 \cite{zhang2022coditt5} by 
19.27\%, 4.82\%, and 5.38\% , respectively, on \sdata in terms of $top\ 1$. \tool relatively improves these three backbone models by 11.36\%, 9.39\%, and 4.49\% on \sdata in terms of $top\ 5$.
On \mdata, \tool improves PLBART \cite{ahmad2021unified}, NatGen \cite{chakraborty2022natgen} and CoditT5 \cite{zhang2022coditt5} by 19.76\%, 2.77\% and 6.32\% relatively in terms of $top\ 1$. 
\tool also improves three backbones by 3.56\%, 5.40\% and 4.44\% on \mdata in terms of $top\ 5$.



\begin{figure}
     \centering
     \begin{subfigure}[b]{0.35\textwidth}
         \centering
         \includegraphics[width=\textwidth]{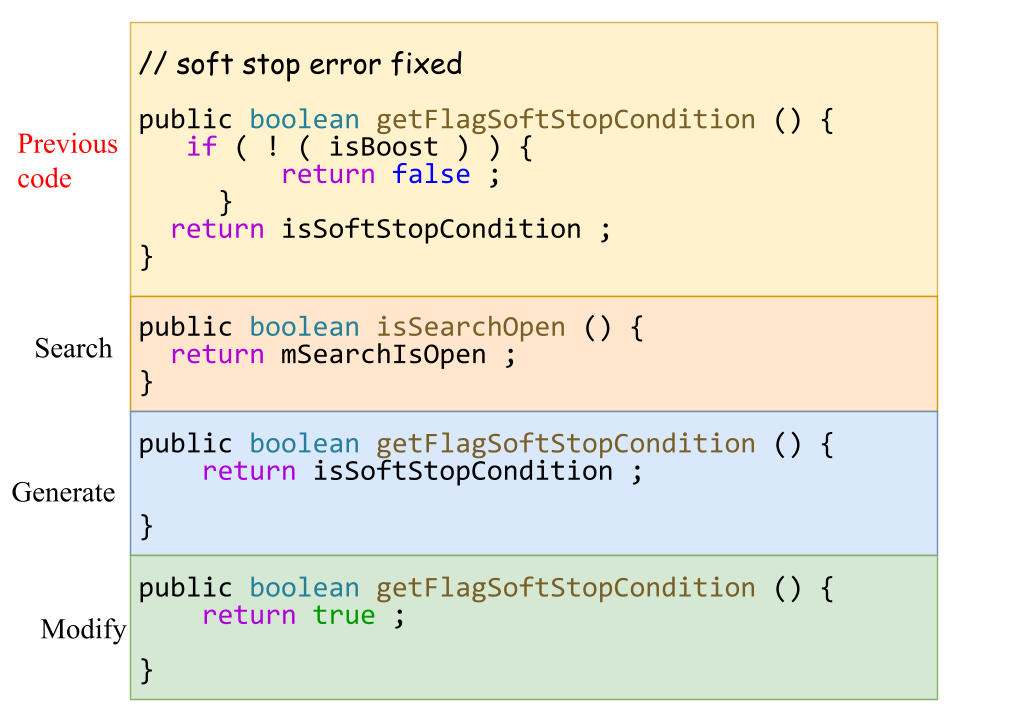}
         \caption{}
         \label{fig:case-study-1}
     \end{subfigure}
     \hfill
     \begin{subfigure}[b]{0.35\textwidth}
         \centering
         \includegraphics[width=\textwidth]{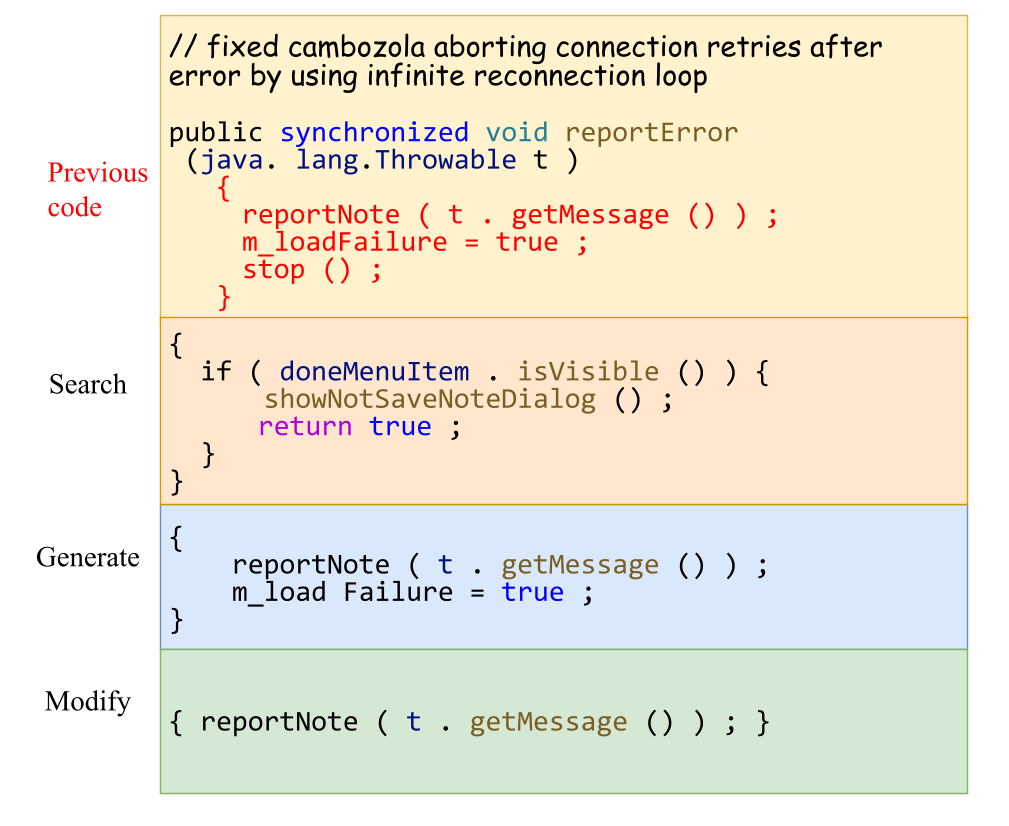}
         \caption{}
         \label{fig:case-study-2}
     \end{subfigure}
     \hfill
        \caption{\small{Example correct patches generated by \tool. Inputs are presented in light brown boxes, and synthesized patches are presented in light green boxes. }}
        \label{fig:rq1-case-study}
\end{figure}

Figure \ref{fig:case-study-1} shows the progress each step makes towards synthesizing the correct patch.
Given the previous code as input, we retrieve a patch that is very similar to the ground truth from the code base. The Levenshtein distance between the  retrieved patch and the ground truth is 2 while that between previous code and ground truth is 14. The generation model (NatGen) utilizes the retrieved patch and generates a patch based on the code context.
This step brings the generated patch one step closer to the correct patch, which is only one step away from our goal.
Finally, the modification model finishes the last step by deleting \texttt{isSoftStopCondition} and inserting \texttt{true}.

Figure \ref{fig:case-study-2} shows another example which can 
prove the robustness of \tool.  ``By using infinite reconnection loop'' in commit message suggests that  \texttt{stop()} should be wiped out from the previous code.
Although the retrieved patch is not even close to the ground truth, 
the generation model (CoditT5 in this case) still recognizes part of the developer's intent and removes \texttt{stop()}. Based on the output of generation model, the editing model further deletes another statement \texttt{m\_loadFailure = true;} and finally returns \texttt{reportNode (t.getMessage()); }, which proves to be the correct patch.


\begin{figure}[h]
     \centering
         \includegraphics[width=0.75\columnwidth]{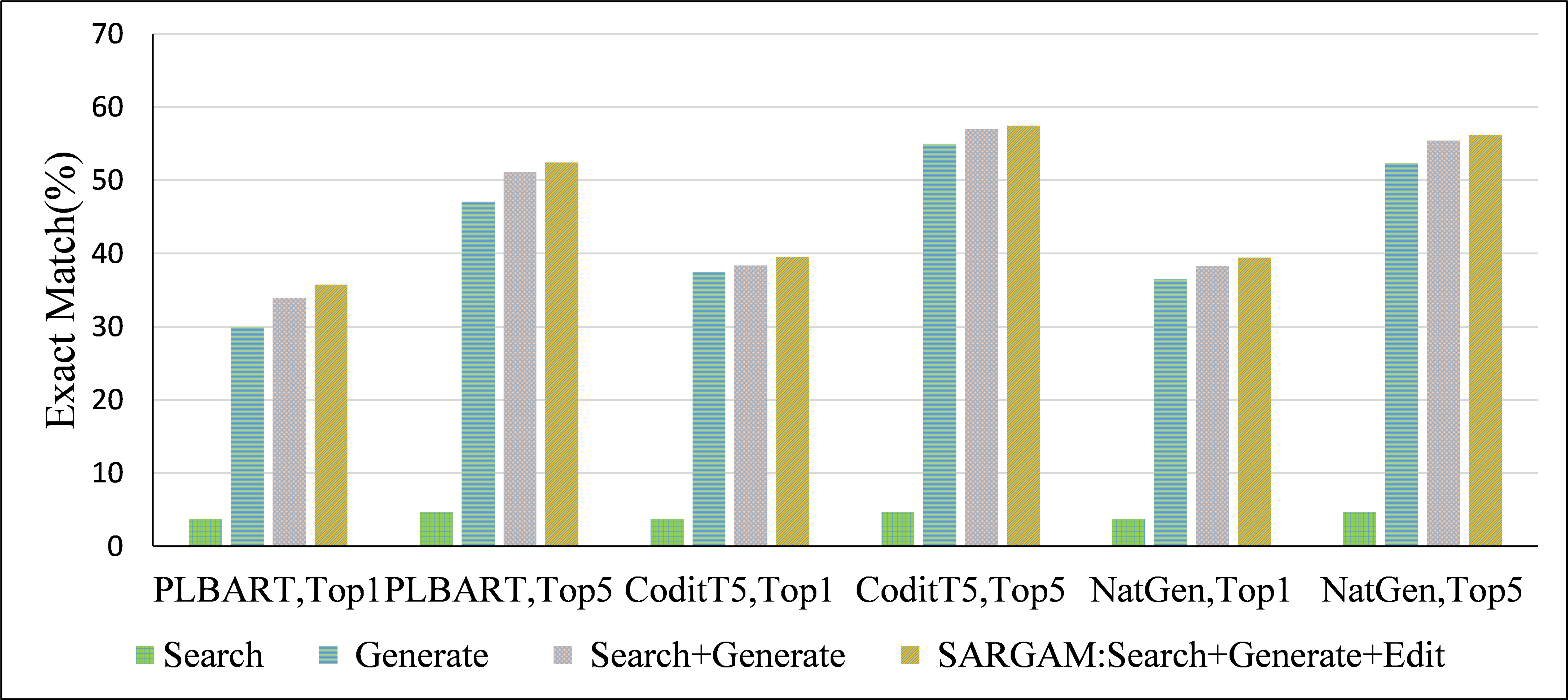}
    \caption{\small{Ablation Study of the steps of \tool on \sdata}}
    \label{fig:ablation_b2f}
\end{figure}
\vspace{-10pt}
Figure \ref{fig:ablation_b2f} further shows the effectiveness of each step (search, generate, modify): for all the three
off-the-shelf code generation models, adding search can improve the patch generation, and modifying the generated patch can further improve the performance. On \sdata retrieved edits can improve the $top1$ exact match of PLBART \cite{ahmad2021unified} by $3.97\%$, and the modifying step further 
improves it with another $1.81\%$. Such an improvement can also be found on \mdata. 



\begin{answerbox}
\textbf{Result 1}:  \tool can generate better patches than generation-only or edit-only models. On average, \tool can improve three baseline models by 8.42\% and 6.44\%, in terms of $top1$ accuracy and $top5$ accuracy, respectively.
\end{answerbox}
\vspace{-20pt}

%% file: body/5e.results_and_analysis_patch_search.tex
\subsection{RQ2. Analyzing Design Choices}

\subsubsection{Motivation}
The success of the search and modification steps depends on different design choices. In particular, we investigate:
\begin{enumerate}
    \item During Search, what is the best method to locate the most similar patch? 
    \item During search, which input modalities (edit location, context, and user intent) and their combinations matter for a successful patch generation? 
    \item How LevT outperform a vanilla Transformer-based model for patch modification?
\end{enumerate}

\subsection*{RQ2.1. Alternatives to locate the most similar patch}
\subsubsection{Experimental Setup} Different combinations of search queries are formed using: patch location, patch context, and developer's intent. Each modality is matched with a similar modality during retrieval. TF-IDF \cite{ramos2003using}, BM25 \cite{robertson2009probabilistic} and PLBART Embedding are used as baselines for patch retrieval. For PLBART Embedding and TF-IDF \cite{ramos2003using}, embeddings of query \query and all instances in database \old are created and cosine similarity is calculated to rank and retrieve the $top\ 5$ similar \old s. Additionally, BM25 algorithm is experimented with to rank and retrieve the $top\ k$ similar \old s. Corresponding patches of the retrieved \old s are fetched. The average Levenshtein distance is computed against the ground truth.

\subsubsection{Results}~\Cref{table:similarity_met} shows the results. 
The average Levenshtein distance is computed and compared against each algorithm. Across all combinations, the best results are achieved when PLBART embedding is used for patch retrieval.

\begin{table}[htbp]
\centering
\vspace{-5pt}
\caption{\small{Impact of Different Similarity Metrics}}
\label{table:similarity_met}
\resizebox{0.80\linewidth}{!}{
\setlength\tabcolsep{1.5pt}
\begin{tabular}{ccc|c|c|c}
\hline
\begin{tabular}[c]{@{}c@{}}Patch \\ Location\end{tabular} & Context & \begin{tabular}[c]{@{}c@{}}Commit \\ Message\end{tabular} & BM25        & PLBART Embedding      & TF-IDF    \\ \hline
 \checkmark                                                         & -       & -                                                         & 0.678 & \textbf {0.633}    & 0.724 \\ \hline
-                                                         &  \checkmark       & -                                                         & 0.719 & \textbf{0.710}    & 0.726 \\ \hline
-                                                         & -       &  \checkmark                                                         & 0.780 & \textbf{0.779}    & 0.780\\ \hline
 \checkmark                                                         &  \checkmark       & -                                                         & 0.703  & \textbf{0.683}   & 0.721        \\ \hline
 \checkmark                                                         & -       &  \checkmark                                                        & 0.692 & \textbf{0.660}  & 0.722        \\ \hline
-                                                         &  \checkmark       &  \checkmark                                                         & 0.724 & \textbf{0.719}   & 0.729        \\ \hline
 \checkmark                                                         & \checkmark       &  \checkmark                                                         & 0.704 & \textbf{0.693} & 0.719        \\ \hline
\end{tabular}

}
\end{table}

\subsection*{RQ2.2. Impact of Input Modalities on Search} 

\subsubsection{Experimental Setup} We form the search query with different combinations of three input types: patch location, patch context, and developer's intent. Each modality is matched with a similar modality during retrieval. We report the results both for search+generate and search+generate+modification as shown in~\Cref{table:varius_retrv}.

\begin{table}[htbp]
\caption{Impact of Different Input Modalities in Search Query on the exact match}
\label{table:varius_retrv}
\begin{center}
\resizebox{0.8\linewidth}{!}%
{
\setlength\tabcolsep{1.5pt}
\begin{tabular}{ccc|rr|rr}
\toprule
  Patch  &           & Commit     &\multicolumn{2}{c|}{Search+Generate}  &\multicolumn{2}{c}{Search+Generate+Modify}\\ 
   Location &   \multirow{-2}{*}{Context} & Message & \sdata & \mdata &\sdata &\mdata \\ 
  \toprule
  - & - &  - &  29.99   & 23.02 &  29.99   & 23.02 \\ \midrule 
  - & - &  \checkmark &  $31.97^\alpha$   & $23.81^\beta$  &$32.02^\alpha$ &$24.26^\beta$\\ \midrule  
  - & \checkmark &  - &  $31.92^\gamma$   & 24.49  &$32.43^\gamma$ &25.92 \\ \midrule 
  - & \checkmark &  \checkmark &  \textbf{33.96}   & 24.43  &35.60 &26.49  \\ \midrule  
  \checkmark & - &  - &  31.50   & 24.72  &32.77 &27.58\\ \midrule 
  \checkmark & - &  \checkmark &  32.82   & \textbf{25.27} &34.01 & \textbf{27.82} \\ \midrule 
  \checkmark & \checkmark &  - &  31.85   & 25.01  &33.29 &26.11\\ \midrule 
  \checkmark & \checkmark &  \checkmark &  33.63   & 24.33 &\textbf{35.77} &26.39 \\ \bottomrule
\end{tabular}
}
\end{center}
\end{table}



\subsubsection{Results}

Table \ref{table:varius_retrv} shows the results of \tool on \sdata and \mdata with different combinations of patches retrieved as an additional modality---the retrieved patches improve the
performance of the generation model, PLBART \cite{ahmad2021unified}, across all combinations.
On \sdata the best result 
is achieved when we use both the patches retrieved with 
context and that retrieved with commit message.
In this case, we improve the performance of PLBART \cite{ahmad2021unified} by 13.24\%.  
However, on \mdata PLBART achieves its best performance
when patches retrieved with patch location and patches retrieved with commit message are passed to the input and it finally improves 
baseline PLBART by 9.77\%. 

The improvement retrieved patches bring to the generation model still holds after further modification. On \sdata, using patches retrieved with all 
the three types of queries achieves the highest accuracy, which is actually the second best combination in the ``Search+Generation" setting. On \mdata, patch location \& commit message is still the best combination.

\begin{table}[htbp]
\centering
\caption{\small{Avg.~Edit Distance Between the Retrieved Patch/Generated Output/Modified Output and the Ground Truth}}

\label{table:averaged_dis}
\resizebox{\linewidth}{!}%
{
\setlength\tabcolsep{1.5pt}
\begin{tabular}{ccc|rr|rr|rr|rr}
\toprule
  Patch  &           & Commit     &\multicolumn{2}{c|}{Before Edit}  &\multicolumn{2}{c}{Search} &\multicolumn{2}{c}{+Generate} &\multicolumn{2}{c}{+Modify}\\ 
   Location &   \multirow{-2}{*}{Context} & Message & \sdata & \mdata &\sdata &\mdata &\sdata &\mdata &\sdata &\mdata\\ 
  \toprule
  - & - &  - &  0.293   & 0.207 &  -   & - &  -   & - &  -   & - \\ \midrule 
  - & - &  \checkmark &  0.293   & 0.207  &0.580 &0.759 &0.236 &0.196 & 0.231 &0.191\\ \midrule  
  - & \checkmark &  - &  0.293   & 0.207  &0.649 &0.701 &0.238 &0.199& 0.234 &0.192\\ \midrule 
  - & \checkmark &  \checkmark &  0.293   & 0.207  &0.652 &0.700 &\textbf{0.232} &0.197 &\textbf{0.225} &0.189 \\ \midrule  
  \checkmark & - &  - &  0.293   & 0.207  &0.579 &\textbf{0.608} &0.240 &0.195 &0.235 &0.190\\ \midrule 
  \checkmark & - &  \checkmark &  0.293   & 0.207 &0.580 & \textbf{0.608} &0.235 &0.196 & 0.231 &\textbf{0.188}\\ \midrule 
  \checkmark & \checkmark &  - &  0.293   & 0.207  &0.581 &0.609 &0.239 &\textbf{0.193}& 0.230 &0.192\\ \midrule 
  \checkmark & \checkmark &  \checkmark &  0.293   & 0.207 &\textbf{0.578} &0.612 & 0.233&0.195 &0.229 &0.192\\ \bottomrule
\end{tabular}
}
\end{table}

Table \ref{table:averaged_dis} also reports the averaged normalized editing distance between the generated patch and the ground truth (GT). Across all combinations, although the retrieved patch is not very similar to GT in terms of normalized editing distance, it is always helping the generation model and the modification model to synthesize patches that are closer to GT.

%% file: body/5d.results_and_analysis_levenshtein.tex
\subsection*{RQ2.3. LevT vs Vanilla Transformer for patch modification}
 
 \subsubsection{Experimental Setup} We follow the setup in \ref{rq1} and use LevT and the vanilla Transformer to modify the output of generation models, which have been augmented with search results. For fairness, both LevT and the vanilla Transformer are trained on the same dataset (\sdata and \mdata).


\begin{table}[]
\caption{\small{Performance (exact match) of LevT and vanilla Transformer (vT) for modification}}
\label{table:rq5}
\centering
\resizebox{0.9\linewidth}{!}{
\begin{tabular}{l|l|ccc|ccc}
\hline
\multirow{2}{*}{\textbf{Dataset}} & \multirow{2}{*}{\textbf{\begin{tabular}[c]{@{}l@{}}Gen.\\ Models\end{tabular}}} & \multicolumn{3}{c|}{\textbf{Top1}}                                                                                                                                  & \multicolumn{3}{c}{\textbf{Top5}}                                                                                                                                   \\ \cline{3-8} 
                                  &                                                                                     &\textbf{\begin{tabular}[c]{@{}c@{}}Before \\ Edit\end{tabular}} & \textbf{\begin{tabular}[c]{@{}c@{}}\\ vT\end{tabular}} & \textbf{\begin{tabular}[c]{@{}c@{}}\\ LevT\end{tabular}} & \textbf{\begin{tabular}[c]{@{}c@{}}Before \\ Edit\end{tabular}} & \textbf{\begin{tabular}[c]{@{}c@{}}\\ vT\end{tabular}} & \textbf{\begin{tabular}[c]{@{}c@{}}\\ LevT\end{tabular}} \\ \hline
\multirow{3}{*}{\textbf{small}}   & PLBART                                                                              & 33.63                & 34.85                                                                    & \textbf{35.78}                                                             & 51.15                & 51.91                                                                    & \textbf{52.43}                                                             \\
                                  & NatGen                                                                              & 38.29                & 39.23                                                                    & \textbf{39.44}                                                             & 56.58                & 57.13                                                                    & \textbf{57.31}                                                             \\
                                  & CoditT5                                                                             & 38.38                & 39.04                                                                    & \textbf{39.55}                                                    & 57.01                & 57.37                                                                    & \textbf{57.47}                                                    \\ \hline
\multirow{3}{*}{\textbf{medium}}  & PLBART                                                                              & 25.27                & 26.32                                                                    & \textbf{27.82}                                                             & 37.57                & 37.98                                                                    & \textbf{38.82}                                                             \\
                                  & NatGen                                                                              & 27.73                & 28.47                                                                    & \textbf{29.32}                                                             & 44.23                & 44.41                                                                    & \textbf{45.31}                                                    \\
                                  & CoditT5                                                                             & 29.29                & 30.03                                                                    & \textbf{30.12}                                                    & 43.53                & 43.68                                                                    & \textbf{44.20}                                                             \\ \hline
\end{tabular}
}
\end{table}

\begin{figure*}
     \centering
     \begin{subfigure}[b]{0.15\textwidth}
         \centering
         \includegraphics[width=\textwidth]{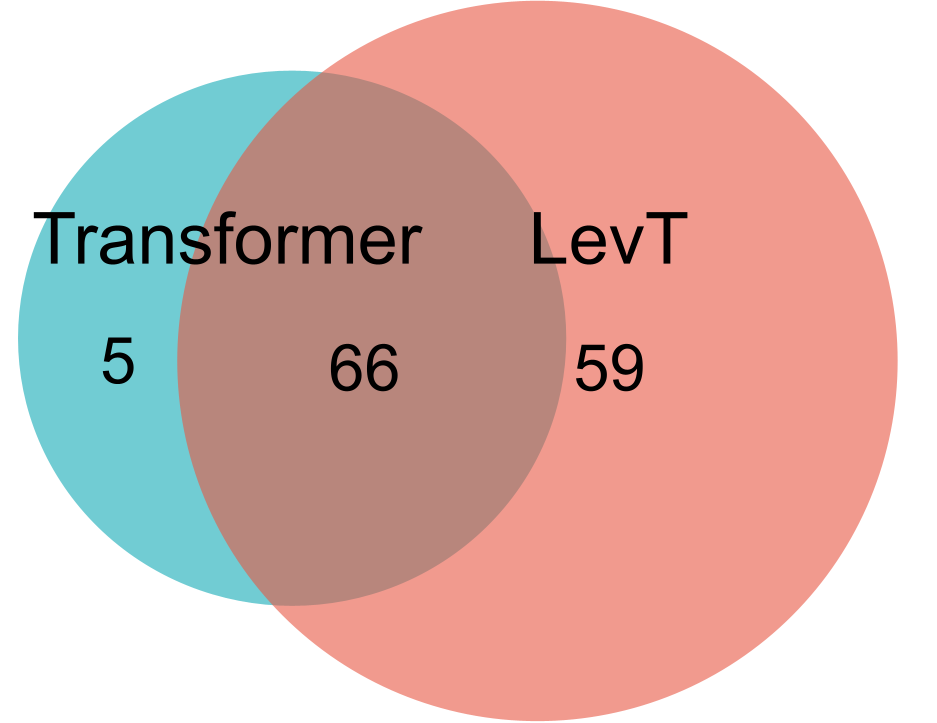}
         \caption{\small{PLBART \sdata}}
         \label{fig:rq5-plbart-small}
     \end{subfigure}
      ~
     \begin{subfigure}[b]{0.15\textwidth}
         \centering
         \includegraphics[width=\textwidth]{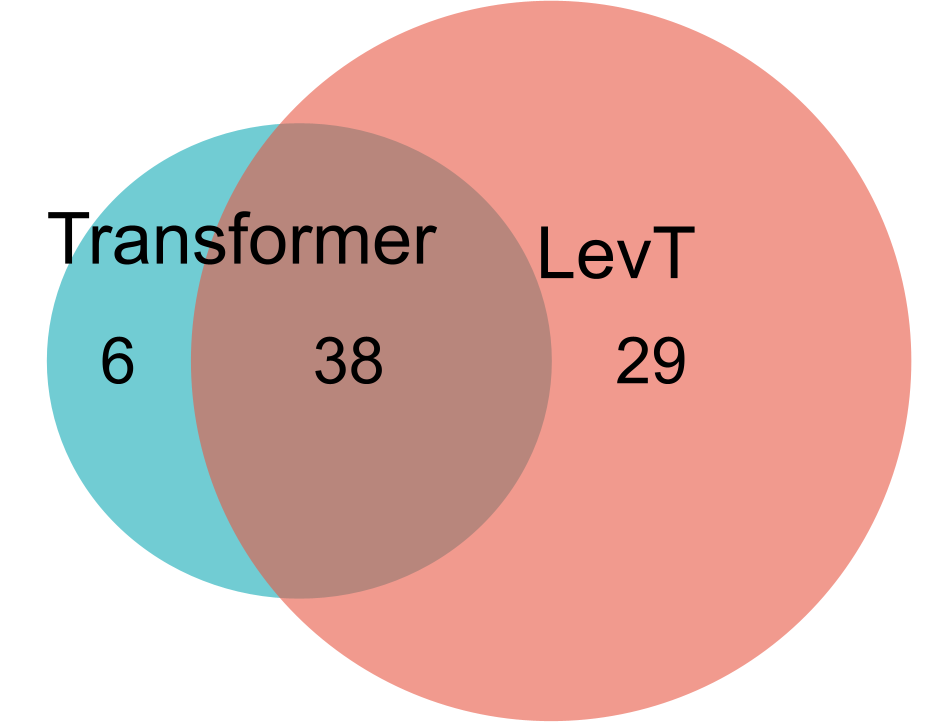}
         \caption{NatGen \sdata}
         \label{fig:rq5-natgen-small}
     \end{subfigure}
       ~
     \begin{subfigure}[b]{0.15\textwidth}
         \centering
         \includegraphics[width=\textwidth]{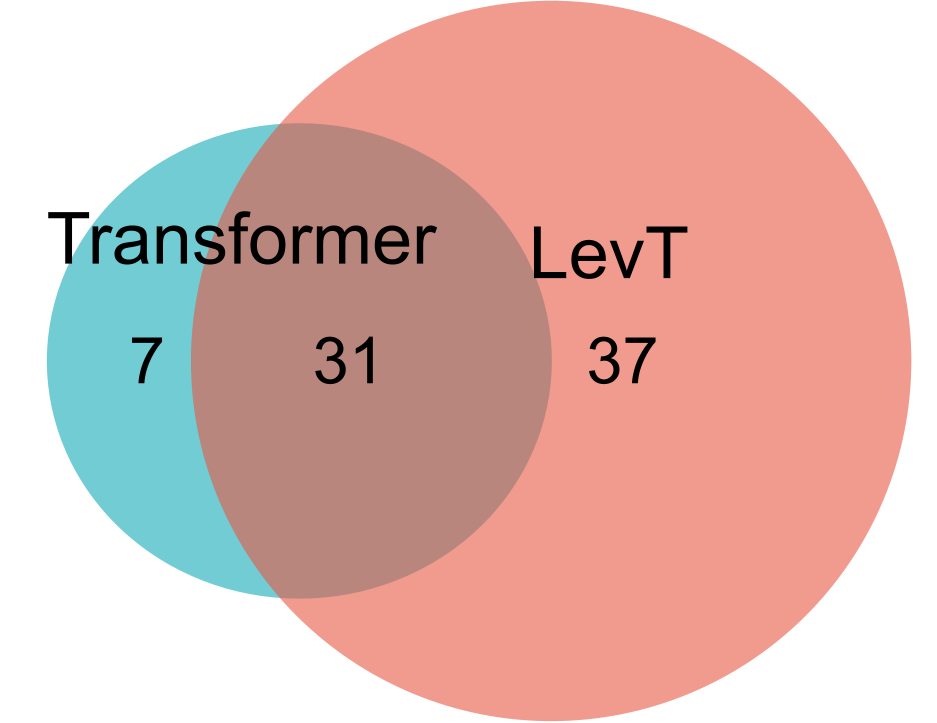}
         \caption{CoditT5 \sdata}
         \label{fig:rq5-coditt5-small}
     \end{subfigure}
     ~
     \begin{subfigure}[b]{0.15\textwidth}
         \centering
         \includegraphics[width=\textwidth]{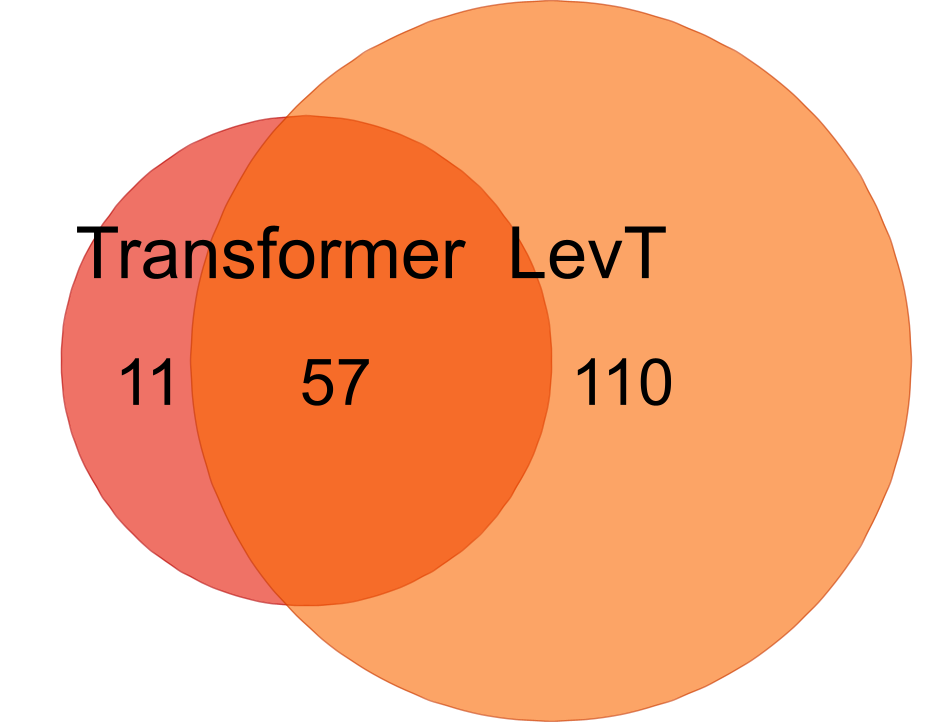}
         \caption{PLBART \mdata}
         \label{fig:rq5-plbart-medium}
     \end{subfigure}
     ~
     \begin{subfigure}[b]{0.15\textwidth}
         \centering
         \includegraphics[width=\textwidth]{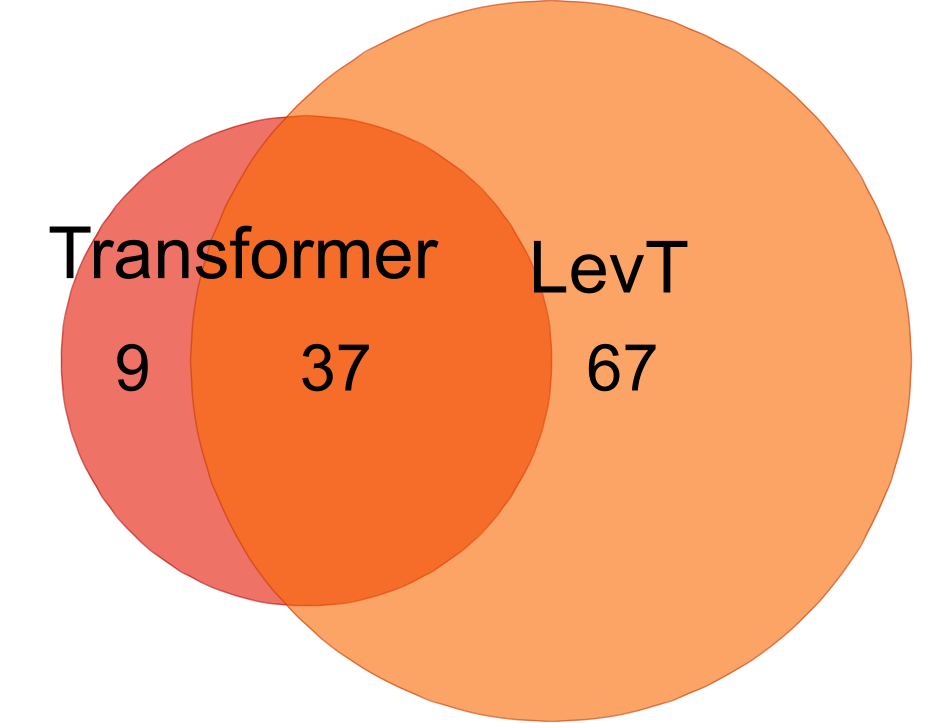}
         \caption{NatGen \mdata}
         \label{fig:rq5-natgen-medium}
     \end{subfigure}
     ~
     \begin{subfigure}[b]{0.15\textwidth}
         \centering
         \includegraphics[width=\textwidth]{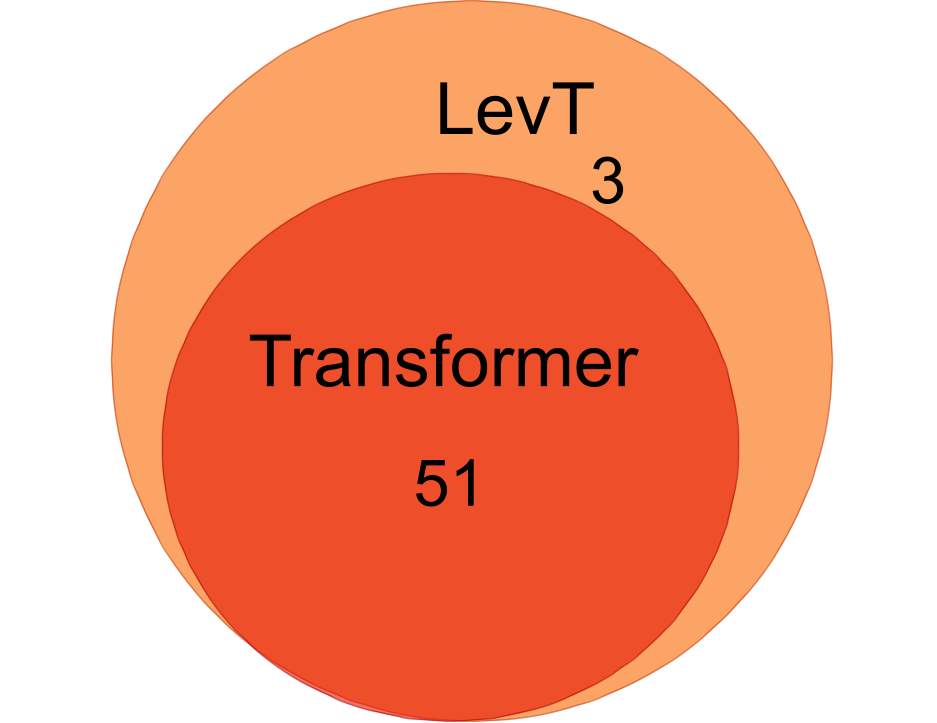}
         \caption{CoditT5 \mdata}
         \label{fig:rq5-coditt5-medium}
     \end{subfigure}
        \caption{\small{Venn diagrams of the numbers of correct modifications made by LevT and Transformer }}
        \label{fig:rq5}
\end{figure*}

 \subsubsection{Results}
 \Cref{table:rq5} reports the performance of using vanilla Transformer and LevT for editing. 
 Across different settings, LevT always achieves a higher exact match (accuracy) of the generated edit.
 In addition, we present the exact 
 numbers of overlapped and unique correct edits produced by Transformer and LevT in Figure \ref{fig:rq5}. On PLBART \mdata and PLBART \sdata, LevT complements Transformer by producing 110 and 59 more correct patches, respectively. Similarly, using by modifying NatGen's output, LevT can produce 29 and 67 more unique patches over vanilla Transformer for \sdata and \mdata respectively. Even when we consider CoditT5, which is an edit generation model, LevT produces 37 and 3 more unique patches over Transformer. These results show LevT is a better design choice for patch modification over vanilla Transformer.

 \begin{answerbox}
\textbf{Result 2} The combinations of edit location, context, and developer's intent during patch retrieval 
can improve PLBART by up to 13.24\%. 
LevT-based patch modification model outperforms the vanilla Transformer 
due to its explicit way of modeling fine granular edit operations. 
\end{answerbox}

%% file: body/5b.results_and_analysis_bug_fixing.tex
\subsection{RQ3. \tool for Bug Fixing}
\label{sec:rq2}
\begin{table}[t]
\caption{Experiment Results (number of correct fixes) of \tool for Bug Fixing. }
\label{table:bug_fixing}
\begin{center}
\resizebox{0.9\columnwidth}{!}{
\begin{tabular}{l|rrr}
\toprule
\textbf{Tool}  & \textbf{Defect4j$_{1.2}$} & \textbf{Defects4j$_{2.0}$} & \textbf{QuixBugs} \\ \toprule
\textbf{CocoNut}                         &    -                   &  -                      & 13           \\
\textbf{CURE}                            &   -                    &    -                    & 26            \\
\textbf{KNOD}                            & 48                  & 34                   & 25              \\
\textbf{AlphaRepair}                     & 45                  & 36                & 28             \\
\textbf{Codex}                           & 33                 & 38                & 31        \\   \midrule
\textbf{\tool}            & 40                 & \textbf{42}                   &    \textbf{34}               \\ 
(Search+Codex+Modify)           &                  &                    &                   \\ 
\bottomrule
\end{tabular}
}
\end{center}
\end{table}
\vspace{-5pt}

\subsubsection{Motivation}
We want to check \tool 's applicability for program repair, which is a special type of code editing task. For bug fixing, the plausibility of the generated edits can be estimated by the passing and failing test cases. 

\subsubsection{Experimental Setup}

Following Jiang et al.'s~\cite{jiang2023impact} findings that Large Language Models (LLM) outperform all other DL-based repair-only models, we choose OpenAI's Codex (at zero-shot setting) \cite{chen2021evaluating}, one of the largest code generation models at the time of writing the paper, for bug fixing. Our goal is to investigate, even using LLM, whether incorporating search and modification steps provides additional benefits.

%

\begin{figure}[h]
\centering
    \includegraphics[width=0.75\linewidth]{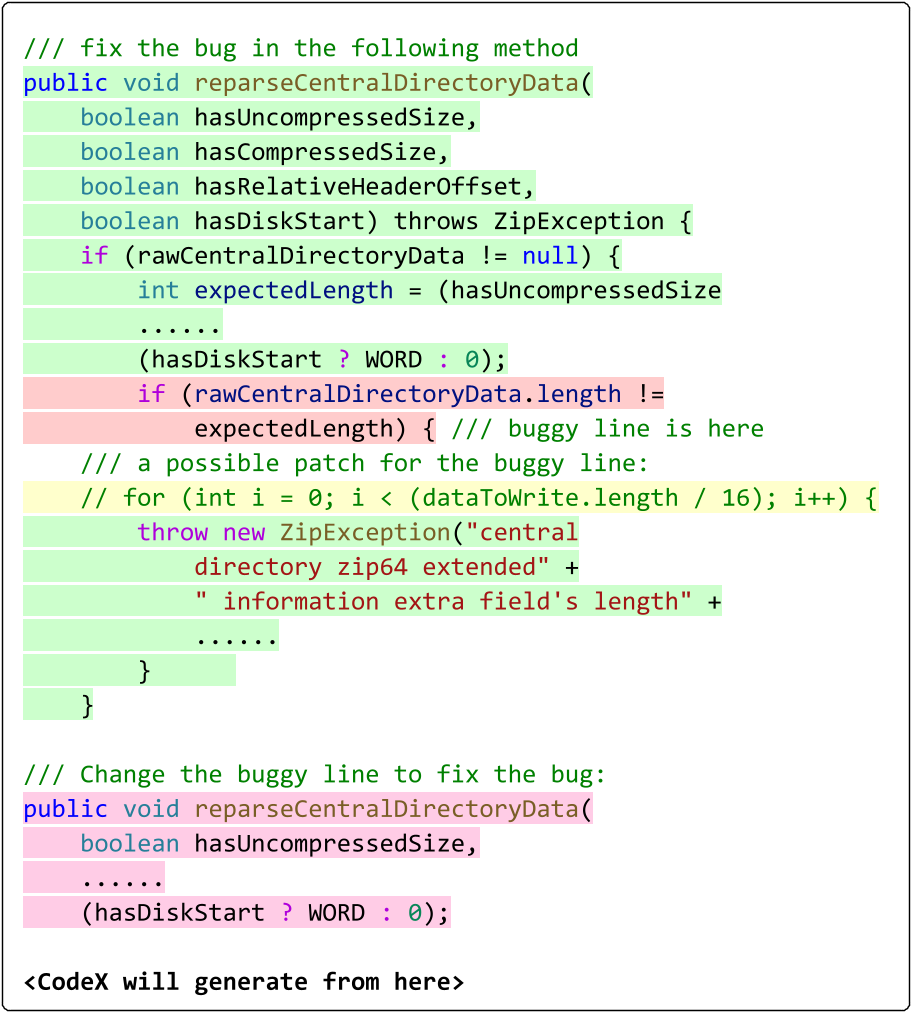}\par
    \centering
    \caption{\small{An example prompt (Codec-17)  
    including the buggy code (green lines), buggy line (red 
    line), retrieved patch (yellow line), and additional context 
    (pink lines). }}
    \label{fig:prompt}
\end{figure}

To provide input to Codex \cite{chen2021evaluating}, we design a prompt that combines code and natural language. Our prompt is inspired by several previous works \cite{prenner2021automatic, joshi2022repair, fan2022improving}. The prompt consists of the following components (see Figure \ref{fig:prompt}):
(i) \textit{Describing Task.} Comment at the beginning of the prompt (\enquote{fix the bug in the following method}) describing Codex's task~\cite{chen2021evaluating};
(ii) The buggy code snippet 
    is marked with a comment \enquote{buggy line is here};
(iii) \textit{Retrieved Patch.} The retrieved patch is augmented with  comment \enquote{A possible patch for buggy line}; and
(iv) \textit{Context.} The context before the buggy line is highlighted with a comment: \enquote{change the buggy line to fix the bug:}.
    
    
    


Here, we perform the search step in a larger training set: Java 2006. In the search step, we retrieve up to 25 similar patches, and in the generation step, we generate top50 possible patches. 
Hence at the inference stage, we obtain up to ($50*25=1250$) candidate patches for every single bug. 
This number of candidate patches is still relatively small compared to the  settings in some previous works \cite{jiang2021cure, xia2022less}, which can generate up to 5,000 patches. 
Here, we use 
 Defects4J test suite to validate patches after each step.
 
 Following previous work \cite{xia2022less}, we call patches
synthesized by \tool 
\enquote{candidate patches}. Then we compile each candidate patch and test it against developer-written test suite to find plausible patches which can pass all the tests. Finally we check if plausible patches are exactly the same as those provided by the developer. Similar to prior work \cite{xia2022less}, we use fix correctness results collected from previous papers. For \dforj and \dforjj, following \cite{jiang2021cure} we only use single-line bugs therefore we filter multiple-lines/hunks bugs out of the results released in the artifacts.  


\subsubsection{Results}
Table \ref{table:bug_fixing} shows the results of \tool and other APR 
baselines on three benchmarks under the condition of perfect bug localization. \tool can fix more bugs than Codex \cite{chen2021evaluating} in all the settings showing that even if we use a really large high-capacity code generation  model, search and modification steps still add values on top of it.

\vspace{3mm}
Overall, \tool fixes 42 single line bugs, outperforming all the other baselines on \dforjj, and produces 6 and 8 more correct patches than the latest APR tools AlphaRepair and KNOD, respectively. 
On \dforj, \tool outperforms most of the deep learning baselines, 
but it is worse than KNOD and AlphaRepair. Note that, we report accuracies based on the top 1250 patches, whereas KNOD and AlphaRepair use 5000 patches. We believe given similar resources we will perform comparably in this setting. 
Table \ref{table:bug_fixing} also presents the effectiveness of the proposed method on QuixBugs where it outperforms all the other baselines.

\begin{figure}
     \centering
     \begin{subfigure}[b]{0.24\textwidth}
         \centering
         \includegraphics[width=\textwidth]{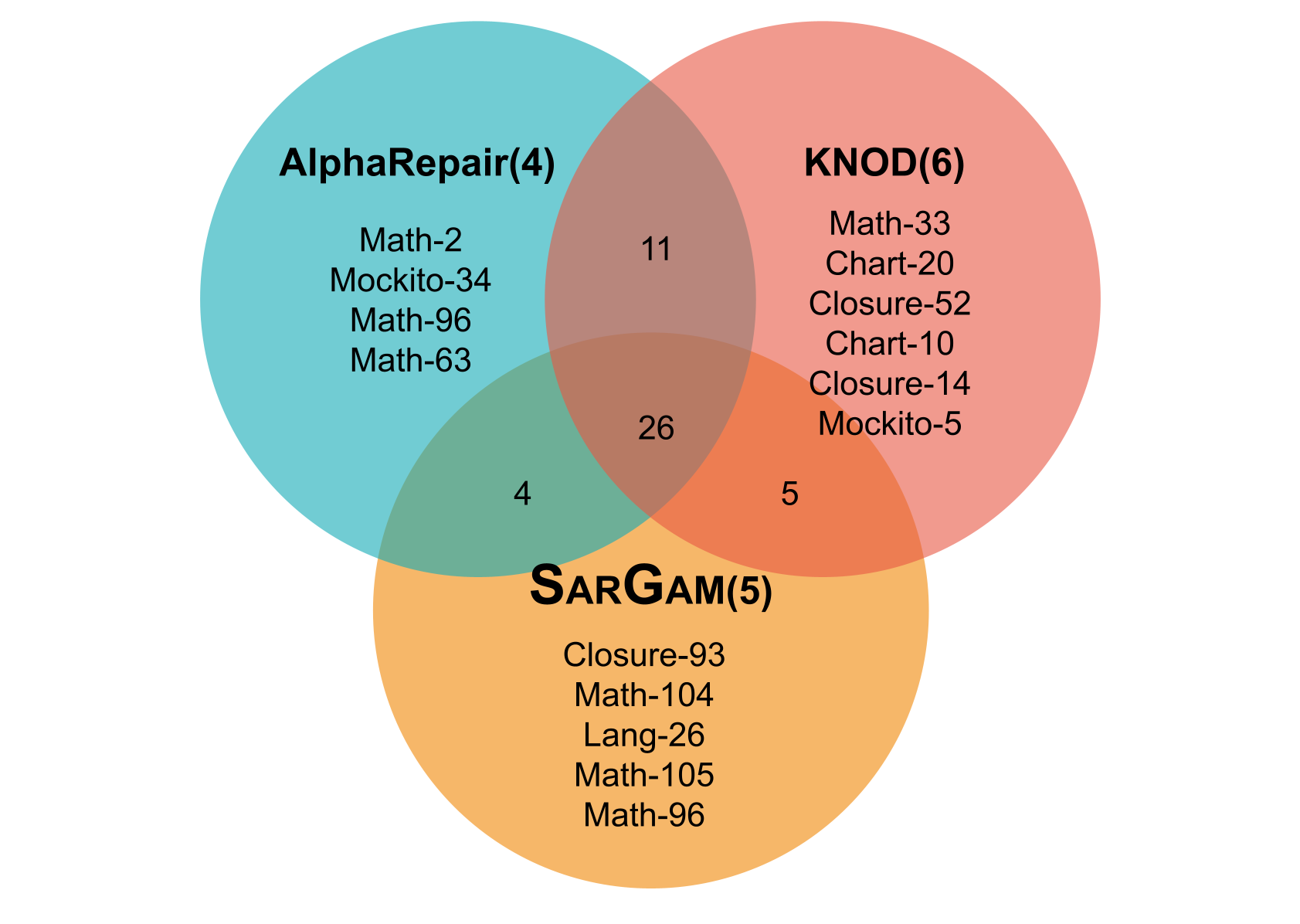}
         \caption{Defects4j$_{1.2}$}
         
         \label{fig:venn_d4j_1}
     \end{subfigure}
     ~
     \begin{subfigure}[b]{0.22\textwidth}
         \centering
         \includegraphics[width=\textwidth]{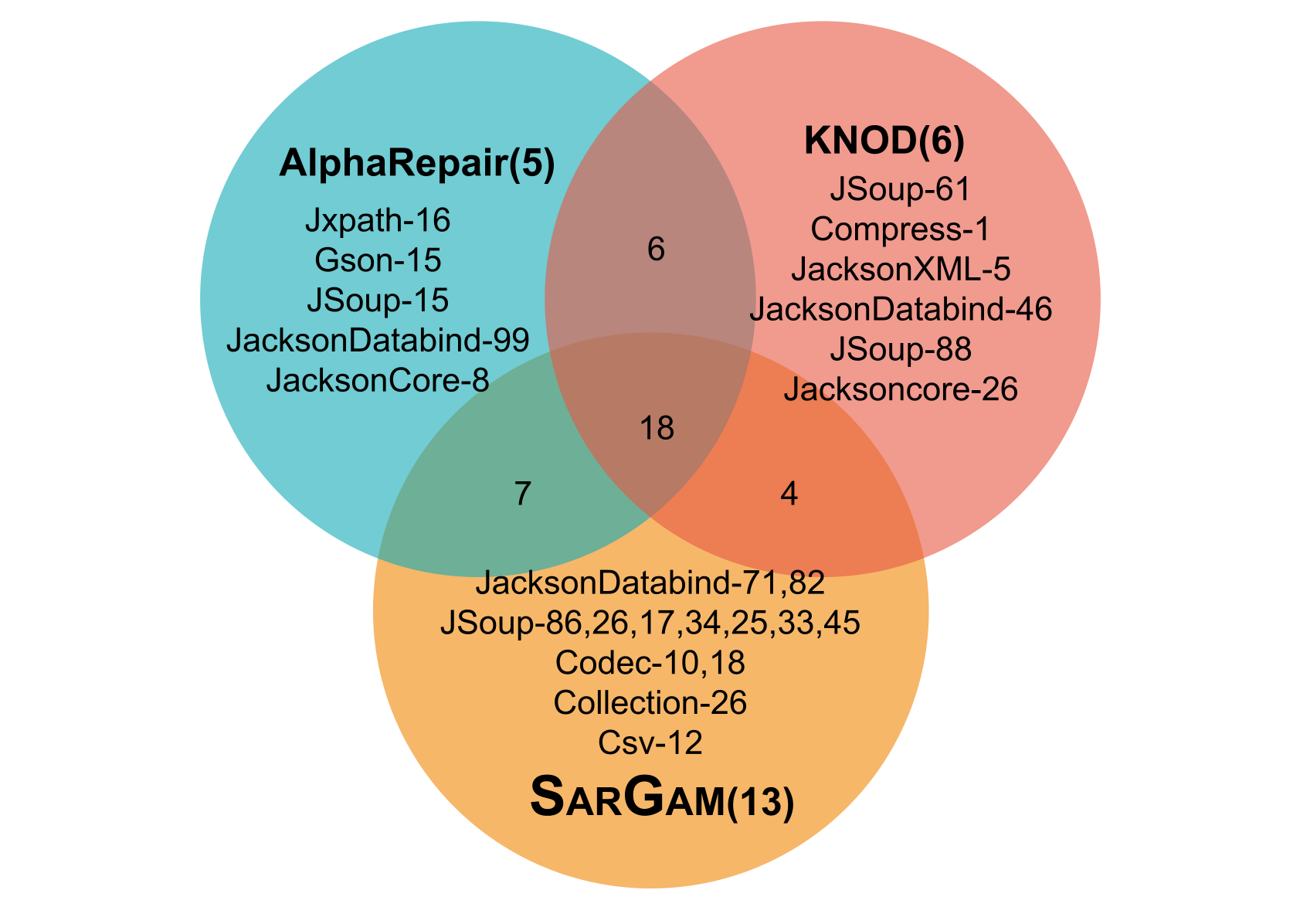}
         \caption{Defects4j$_{2.0}$}
         \label{fig:venn_d4j_2}
     \end{subfigure}
        \caption{\small{Unique fixes of \tool, AlphaRepair and KNOD. }}
        \label{fig:venn_d4j}
\end{figure}

\begin{figure}
\centering
     \begin{subfigure}[b]{0.4\textwidth}
         \centering
\includegraphics[width=\textwidth]{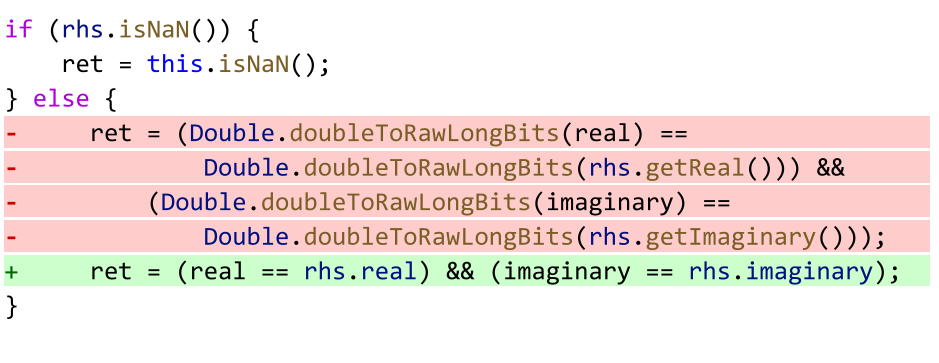}
         \caption{\dforj Math-96}
         \label{fig:math-96}
     \end{subfigure}
     \hfill
     \begin{subfigure}[b]{0.4\textwidth}
         \centering
         \includegraphics[width=\textwidth]{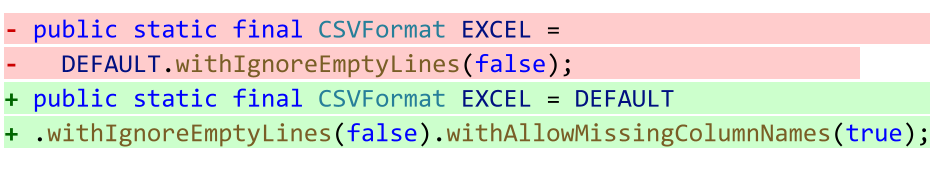}
         \caption{\dforjj Csv-12}
         \label{fig:csv-12}
     \end{subfigure}
     \hfill
     \begin{subfigure}[b]{0.4\textwidth}
         \centering
         \includegraphics[width=\textwidth]{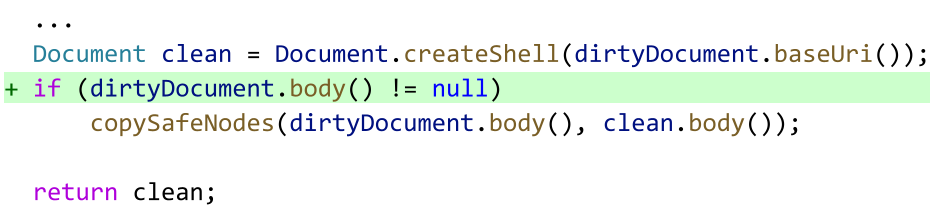}
         \caption{\dforjj JSoup-26}
         \label{fig:jsoup-26}
     \end{subfigure}
        \caption{\small{Unique bugs only fixed by \tool}}
        \label{fig:unique_bugs}
\end{figure}

Figure \ref{fig:venn_d4j} demonstrates \tool's unique bug-fixing capabilities alongside AlphaRepair and KNOD, with Figure \ref{fig:venn_d4j_1} showing \tool fixing additional bugs on \dforj and \dforjj—10 and 17 more than AlphaRepair, and 9 and 20 more than KNOD, respectively. Figure \ref{fig:venn_d4j} demonstrates \tool's unique bug-fixing capabilities. 
Math-96 (Figure \ref{fig:math-96}) is a hard bug because all
the  {\tt Double.\-doubleTo\-RawLongBits} need to be deleted from the original sequence. 
Csv-12(\ref{fig:csv-12}) is also nontrivial because a new api method {\tt .withAllowMissingColumnNames(true)} is called in the correct fix and it does not appear in the context. However, \tool is still able to fix both of them with the help of patch search and patch editing. Another example is JSoup-26 (Figure \ref{fig:jsoup-26}), which indicates that \tool is able to insert a new line into the buggy code.

\begin{answerbox}
\textbf{Result 3}: \tool is capable of fixing bugs---on the real-world Java bug dataset, \tool can synthesize 6 and 8 more bugs than most recent AlphaRepair and KNOD.
\end{answerbox}
\vspace{-12pt}

%% file: body/6.related_work.tex
\section{Related Work}
\textbf{Code Repair Models. } Seq2Seq models are widely explored for APR. Tufano et al. \cite{tufano2019empirical} applied the encoder-decoder model for bug fixing. 
SequenceR \cite{chen2019sequencer} enhanced Seq2Seq models with a copy mechanism to address the vocabulary issue. 
CocoNut \cite{lutellier2020coconut} employed ensemble learning. Additionally, some research \cite{chakraborty2020codit, jiang2023knod, dinella2019hoppity} has adopted tree/graph-based models.

LLMs like CodeBERT \cite{feng2020codebert}, PLBART \cite{ahmad2021unified}, CodeT5 \cite{wang2021codet5}, and NatGen \cite{chakraborty2022natgen} have demonstrated significant success in APR. 
VRepair \cite{chen2021neural} and VulRepair \cite{fu2022vulrepair} are T5-based model to repair vulnerabilities. 
Recent studies explore LLMs for zero-shot APR, eliminating the need for additional training or fine-tuning~\cite{jiang2023impact}. 
Xia et al. \cite{xia2022less} used pre-trained CodeBERT for a cloze-style APR tool. 
Other works \cite{prenner2021automatic, joshi2022repair} crafted prompts for Codex \cite{chen2021evaluating} for code repair as a code generation task.

The plastic surgery hypothesis \cite{barr2014plastic} suggests that codebase changes often reuse existing snippets, which can be effectively identified and utilized. 
Despite this, many current APR methods, as derived from Neural Machine Translation, overlook leveraging the evolutionary codebase. 
Our approach enhances this framework by demonstrating how integrating retrieval steps with standard APR tools can unlock additional potential.

\textbf{Retrieval-based Code Repair Models.}
Previous work has focused on reusing code for bug repair. Xin et. al.~\cite{xin2017leveraging, xin2019revisiting} search a code database for code snippets similar to the bug context and reuses them to synthesize patches. 
LSRepair~\cite{liu2018lsrepair} suggests that code search accelerates the repair process, fixing some bugs in seconds. 
These studies inspire us to investigate whether patches from the codebase can enhance $seq2seq$ models and boost their performance. 
However, they depend heavily on the codebase quality; We overcome these limitations by combining code search with generation and modification models. 
These two components have the capability to adaptively leverage the useful information from the retrieval results and creatively synthesis patching patterns that are never seen in the existed codebase. 



\textbf{Code Editing Models}
Recent studies investigate DL models' ability to learn explicit edit operation~\cite{ding2020patching}. Chen et al. \cite{chen2021plur} introduced pairing a graph encoder with a Transformer decoder to produce Tocopo sequences \cite{tarlow2020learning}, representing code edits. Zhang et al. \cite{zhang2022coditt5} developed CoditT5, a pre-trained model tailored for editing tasks. Differently, \tool doesn't directly produce an edit sequence but progressively steers the tool to generate the intended edit through multiple steps.~\cite{tarlow2020learning,chen2021plur,connorcan2021,zhang2022coditt5} developed specialized outputs for edit operations. CoditT5 \cite{zhang2022coditt5} generates an edit plan outlining explicit operations before producing the edited target sequence, necessitating additional post-processing. Unlike these approaches, LevT directly incorporates explicit edit operations into the decoder, bypassing the need for specialized output designs.

%% file: body/7.threats.tex
\vspace{-12pt}
\section{Threats To Validity}

\noindent
\textit{External Validity.} Our edit generation evaluation relies on two datasets, \sdata and \mdata, focusing on smaller edits and possibly missing broader edit characteristics. Commit messages used as edit intents can also be noisy. 
Despite these limitations, these datasets reflect real development practices. We also test \tool on three additional datasets common in APR research, ensuring our findings are robust and broadly applicable across different edit generation scenarios.


\noindent
\textit{Construct Validity.} \tool needs precise edit locations for input.  While developers' cursors can simulate the edit location during edit generation. However, determining the exact location of a bug can be more challenging; We use other tools for bug detection, allowing \tool to focus on generating high-quality patches once the bug is pinpointed. This strategy emphasizes \tool's patch generation capabilities and avoids problems related to incorrect bug location.


\noindent
\textit{Internal Validity.} Our results may be influenced by our choices of hyperparameters used in the model (\eg learning rate, batch size, etc.). To address this concern, we released our tool as an open-source project so that other researchers and practitioners will be able to evaluate our approach in a wider range of settings, which will help to validate our findings further and minimize this potential threat. 

%% file: body/7.5.future_work.tex
\vspace{-12pt}
\section{Discussion \& Future work}

\textbf{Discussion.}
The primary technical innovation in our approach is the introduction of a new edit model utilizing the Levenstein Transformer. Unlike code generation, code editing incorporates the likelihood of generating edits based on a prior version of the code. To the best of our knowledge, the field of code editing models remains relatively unexplored in software engineering research, with only one previous study by Zhang et al. (2022) directly addressing edit modeling. Our empirical results demonstrate that our model surpasses the previous work, achieving improvements of 2.02\% and 2.47\% in Top1 and Top5 accuracy, respectively, on the \sdata dataset.


Further, we systematically capture developers' code editing behaviors within a unified framework, and empirically demonstrate:
(i) Retrieval-based techniques can help a vanilla edit generation-based technique and vice versa (Table \ref{table:varius_retrv} and Table \ref{table:averaged_dis}); (ii) A modification model can be very effective even after retrieval augmented generation; and (iii) We propose and implement a new pipeline to combine the above three steps together and prove their effectiveness.


We aim to extend our current work as follows.

\textbf{Real World Evaluation.} User study is always a crucial way to evaluate the effectiveness of tools. 
We plan to evaluate the utility of \tool through a questionnaire and ask developers about their opinion after their actual use of \tool.

\edit{\textbf{Limited Input Window.}} The representation of the input can play an important role for Transformer models. For code editing task, we follow the input representation of 
\cite{chakraborty2021multi, zhang2022coditt5}, which first concatenate all the modalities and then truncate from behind if the length of the sequence exceeds the window size. \edit{ However, we may lose parts of the retrieved patch if the original patch location and its patch context are too long. Another option could be to set a window size for each of the modalities and apply truncation separately.} This can ensure that all the modalities are preserved even after truncation.

\edit{\textbf{Brute-force Search}. Our patch search method employs a brute-force approach by sequentially examining code samples. 
This method may struggle to scale for large datasets, indicating a need for optimization towards a more efficient algorithm.}



%% file: body/8.conclusion.tex
\vspace{-15pt}
\section{Conclusion}
We propose \tool, a novel approach to improve pre-trained code generation models by incorporating patch search \& retrieval and patch modification. Our goal is to mimic the behavior of a developer while editing, who first searches for a related patch, writes a sketch, and then modifies it accordingly. To this end, we propose a novel patch modification model based on Levenshtein Transformer, 
which generates fine-granular edit operations to realize patch modification. We evaluate our approach on two tasks: code editing and automated program repair. Our results demonstrate that \tool is highly effective for both tasks and outperforms state-of-the-art methods in most settings.